# Electromagnetic force and torque derived from a Lagrangian in conjunction with the Maxwell-Lorentz equations

Masud Mansuripur

James C. Wyant College of Optical Sciences, The University of Arizona, Tucson, Arizona 85721

**Abstract**. Electromagnetic force and torque are typically derived from a stress tensor in conjunction with Maxwell's equations of classical electrodynamics. In some instances, the Principle of Least Action (built around a Lagrangian) can be used to arrive at the same mathematical expressions of force and torque as those derived from a stress tensor. This paper describes some of the underlying arguments for the existence of a Lagrangian in the case of certain simple physical systems. While some formulations of electromagnetic force and torque admit a Lagrangian, there are other formulations for which a Lagrangian may not exist.

**1. Introduction**. The principle of least action is a powerful tool in classical physics that enables one to arrive at the equations of motion of a physical system by finding the path, corresponding to a local minimum (or local extremum) of the action, that takes the system from a fixed initial position (in spacetime or in coordinate space) to a fixed final position.[1-3] The action itself is the integral of a Lagrangian taken from the initial time $t_0$ to the final time $t_1$ along any path that the system might take in going from an initial location to a final location in its coordinate space. The theoretical methods of analysis based on the Lagrangian, a function of the coordinates and their corresponding velocities (and also those based on a related function, the Hamiltonian) are indispensable in modern physics, in general, and in quantum mechanics and quantum electrodynamics, in particular.[4]

The goal of the present paper is to discuss the equations of motion of a point-particle of mass $m$ and electric charge $q$ that, in addition, possesses an electric dipole moment $\boldsymbol{p}(t)$, in the presence of external electromagnetic (EM) fields. Starting with a simple Lagrangian that takes into account the existence of a magnetic dipole moment $\boldsymbol{m}(t)$, which inevitably accompanies a moving electric dipole,[5-8] we derive the expressions of EM force and torque acting on the particle. We demonstrate the consistency of the force and torque formulas derived from the postulated Lagrangian with those obtained directly from the Lorentz force law.[3-7] The Lorentz force acting on the dipole pair will be seen to include a contribution from the so-called "hidden momentum," originating in the interaction between the external electric field $\boldsymbol{E}(\boldsymbol{r},t)$ and the magnetic dipole moment $\boldsymbol{m}(t)$ moving along the particle trajectory $\boldsymbol{r}_p(t)$.[9-17] When the hidden momentum contribution is removed from the expression of the EM force on the point-particle, the result turns out to be the Einstein-Laub force.[18]

It will be seen that the Lorentz torque acting on the accompanying magnetic dipole $\boldsymbol{m}(t)$ differs from what might naively be expected based on the similarities between electric and magnetic dipoles. We proceed to show that the simple Lagrangian that works so well for a moving electric dipole (accompanied by a relativistically-induced magnetic counterpart), fails to yield the equations of motion for a moving magnetic dipole — accompanied, as it must, by an electric dipole counterpart — in the presence of external EM fields. The reasons for this failure are also discussed.

Previously published work on the subject includes discussions of EM force and/or torque on stationary as well as mobile dipoles within uniform fields, static fields, or dynamic fields.[19-22] The reader should consult the cited references (and the references cited therein) for more information.

The organization of the paper is as follows. In Sec. 2, we derive the equation of motion for a point-charge traveling within an external EM field. Here, the standard Lagrangian, expressed in terms of the scalar and vector potentials of the EM field, is used to show that the time-rate-of-change of the (relativistic) momentum of a point-particle of mass $m$ and charge $q$ equals the Lorentz force exerted by the $\boldsymbol{E}$ and $\boldsymbol{B}$ fields.

Section 3 is devoted to a derivation of the equation of motion of a point-charge $q$ accompanied by an electric point-dipole $\boldsymbol{p}(t)$ in an external EM field. Given that a moving electric dipole is always accompanied by a relativistically-induced magnetic dipole $\boldsymbol{m}(t)$,[5-8] the Lagrangian in this case contains two additional terms corresponding to the potential energies of the conjoined dipole pair. (A justification for the existence of the accompanying magnetic dipole is given in Appendix A, followed by a derivation, in Appendix B, of the Lagrangian for the dipole pair starting from their electric charge and current densities.) The Lagrangian introduced in Sec. 3 is subsequently used in Sec. 4 to arrive at a relativistic expression for the Hamiltonian of the point-particle [i.e., the particle of mass $m$, charge $q$, electric dipole moment $\boldsymbol{p}(t)$, and the induced magnetic dipole moment $\boldsymbol{m}(t)$] in the presence of an external EM field.[4]

In Section 5, the Lagrangian of the dipole pair is augmented by the particle's rotational kinetic energy, thus enabling us to obtain the dipole pair's equations of rotary motion under the influence of external EM fields. In this way, the EM torque acting on a moving electric dipole is derived from the postulated Lagrangian. We confirm that the EM force and torque acting on the moving dipole as derived from the postulated Lagrangian are consistent with the expressions of force and torque that one can directly derive from the Lorentz force law; this is done in Appendix C for the force formula, and in Appendix D for the torque formula. A major difference between the Lorentz force and the Einstein-Laub force is due to a contribution by the hidden momentum, which is manifest in all our calculations.[23] (The hidden momentum arises when an electric field acts on a magnetic dipole, a condition that is patently present in the problem under consideration, given that the motion of our electric dipole automatically gives rise to an accompanying magnetic dipole.) In Appendix E, we verify, once again, the contribution of the hidden momentum to the Lorentz force by starting with the Einstein-Laub force acting on the dipole pair, and proceeding to show that it leads to the same formula as that of the Lorentz force, except for an additional term that is the time-rate-of-change of the hidden momentum.

In an attempt to extend the same Lagrangian formalism to a moving magnetic dipole (accompanied by its relativistically-induced electric dipole counterpart[5-7]), we ran into difficulties stemming from a certain lack of symmetry between this and the case of a moving electric dipole. A detailed discussion of these difficulties is relegated to Appendix F, where we elucidate the fundamental difference between the Lagrangian of a moving electric dipole and that of a moving magnetic dipole. Our inability to apply the aforementioned Lagrangian formalism to a moving magnetic dipole does *not* prevent us from directly computing the Lorentz force, or the Einstein-Laub force, experienced by such a dipole (along with its accompanying electric dipole counterpart) within an external EM field. The latter part of Appendix F is devoted to computing the Einstein-Laub force on a moving magnetic dipole, which, aside from the hidden momentum term, turns out to be identical to the Lorentz force—a behavior that precisely parallels that of a moving electric dipole described in Appendix E.

The mathematical derivations in the main body of the paper are given in some detail, lest the vector-algebraic subtleties obscure the logical arguments behind the formulas. The derivations in the appendices, however, rely on the full machinery of vector calculus and, as such, they are more compact and contain several shortcuts. The paper closes in Sec. 6 with a brief summary and a few final remarks.

**2. A point-charge in an external electromagnetic field**. The relativistic Lagrangian for a point-particle of charge $q$ and mass $m$, moving with velocity $\boldsymbol{V}(t)$ in an EM field specified by the scalar and vector potentials, $\psi(\boldsymbol{r},t)$ and $\boldsymbol{A}(\boldsymbol{r},t)$, is given by[1-4]

$$\mathcal{L}(r, V, t) = -mc^2\sqrt{1 - (V/c)^2} - q\psi(r, t) + qV \cdot A(r, t). \tag{1}$$

The particle's position and velocity are $r(t) = x(t)\hat{x} + y(t)\hat{y} + z(t)\hat{z}$ and $V(t) = \dot{r}(t)$. In what follows, we shall use $V$ and $\dot{r}(t)$ interchangeably. The equation of motion of the point-particle is derived straightforwardly from the above Lagrangian, as follows:

$$\frac{\partial \mathcal{L}}{\partial \dot{x}} = \frac{mV_x}{\sqrt{1-(V/c)^2}} + qA_x(r, t). \tag{2a}$$

$$\frac{d}{dt}\left(\frac{\partial \mathcal{L}}{\partial \dot{x}}\right) = \frac{d}{dt}\left(\frac{mV_x}{\sqrt{1-(V/c)^2}}\right) + q\left(\frac{\partial A_x}{\partial t} + \frac{\partial A_x}{\partial x}V_x + \frac{\partial A_x}{\partial y}V_y + \frac{\partial A_x}{\partial z}V_z\right). \tag{2b}$$

$$\frac{\partial \mathcal{L}}{\partial x} = -q\frac{\partial \psi}{\partial x} + q\left(V_x\frac{\partial A_x}{\partial x} + V_y\frac{\partial A_y}{\partial x} + V_z\frac{\partial A_z}{\partial x}\right). \tag{2c}$$

Defining the particle's (relativistic) kinetic momentum as $p = mV/\sqrt{1 - (V/c)^2}$, then equating Eq.(2b) with Eq.(2c), yields

$$\frac{d}{dt}\left(\frac{\partial \mathcal{L}}{\partial \dot{x}}\right) = \frac{\partial \mathcal{L}}{\partial x} \rightarrow \frac{dp_x}{dt} = -q\left(\frac{\partial \psi}{\partial x} + \frac{\partial A_x}{\partial t}\right) + qV_y\left(\frac{\partial A_y}{\partial x} - \frac{\partial A_x}{\partial y}\right) - qV_z\left(\frac{\partial A_x}{\partial z} - \frac{\partial A_z}{\partial x}\right)$$

$$\rightarrow \frac{dp_x}{dt} = qE_x(r, t) + qV_yB_z(r, t) - qV_zB_y(r, t) \rightarrow \frac{dp_x}{dt} = qE_x(r, t) + q(V \times B)_x. \tag{3}$$

As usual, the EM fields are given by $E(r, t) = -\nabla\psi - \partial A/\partial t$ and $B(r, t) = \nabla \times A(r, t)$.[2-7] The equations of motion along the $y$ and $z$ directions are found similarly to that along the $x$-axis as given by Eq.(3). It is thus clear that the standard Lagrangian of Eq.(1) leads to the relativistic version of Newton's second law, $F = \dot{p}(t)$, with the local and instantaneous force acting on the particle being the Lorentz force $F = q(E + V \times B)$.

The canonical momentum $P$ of the point-particle may now be obtained from the identity $P_x = \partial \mathcal{L}/\partial \dot{x}$ using Eq.(2a), as follows:[4]

$$P_x = \partial \mathcal{L}/\partial \dot{x} = \left[mV_x/\sqrt{1 - (V/c)^2}\right] + qA_x \rightarrow \underbrace{P(r, t)}_{\text{canonical momentum}} = \underbrace{p(r, t)}_{\text{kinetic momentum}} + qA(r, t). \tag{4}$$

The above equation yields the following expressions for the velocity $V$ and the relativistic factor $\sqrt{1 - (V/c)^2}$:

$$\frac{mV}{\sqrt{1 - (V/c)^2}} = P - qA \rightarrow \frac{m^2V^2}{1 - (V/c)^2} = (P - qA)^2 \rightarrow \frac{1}{1 - (V/c)^2} = 1 + \frac{(P-qA)^2}{(mc)^2}. \tag{5}$$

The Hamiltonian of the point-particle may now be written as follows:[4]

$$\mathcal{H}(r, P, t) = \sum_\ell P_\ell \dot{x}_\ell - \mathcal{L}(r, \dot{r}, t) = P \cdot V + mc^2\sqrt{1 - (V/c)^2} + q\psi(r, t) - qV \cdot A(r, t)$$

$$= mc^2\sqrt{1 - (V/c)^2} + (P - qA) \cdot V + q\psi(r, t)$$

$$= mc^2\sqrt{1 - (V/c)^2} + m^{-1}(P - qA)^2\sqrt{1 - (V/c)^2} + q\psi(r, t)$$

$$= mc^2\sqrt{1 - (V/c)^2}\left[1 + (P - qA)^2/(mc)^2\right] + q\psi(r, t)$$

$$= mc^2[1 + (P - qA)^2/(mc)^2]^{1/2} + q\psi(r, t). \tag{6}$$

The non-relativistic version of the above Hamiltonian for $V \ll c$ is obtained by retaining only the first two terms in the Taylor series expansion of the square root; that is,

$$\mathcal{H}(\mathbf{r}, \mathbf{P}, t) \cong mc^2 + [(\mathbf{P} - q\mathbf{A})^2/2m] + q\psi(\mathbf{r}, t). \tag{7}$$

As expected, the non-relativistic Hamiltonian of the point-charge turns out to be the sum of its rest energy $mc^2$, kinetic energy $\mathbf{p}^2/2m$, and the (scalar) potential energy $q\psi(\mathbf{r}, t)$.

**3. An electric point-dipole moving within an external EM field.** The relativistic Lagrangian for a point-particle of charge $q$, mass $m$, and electric dipole moment $\boldsymbol{p}(t)$, moving with velocity $\mathbf{V}(t)$ in an external EM field specified by the scalar and vector potentials $\psi(\mathbf{r}, t)$ and $\mathbf{A}(\mathbf{r}, t)$, which are associated with the electric and magnetic fields $\mathbf{E}(\mathbf{r}, t)$ and $\mathbf{B}(\mathbf{r}, t) = \mu_0 \mathbf{H}(\mathbf{r}, t)$, is given by

$$\mathcal{L}(\mathbf{r}, \mathbf{V}, t) = -mc^2\sqrt{1 - (V/c)^2} - q\psi(\mathbf{r}, t) + q\mathbf{V} \cdot \mathbf{A}(\mathbf{r}, t) + \boldsymbol{p}(t) \cdot \mathbf{E}(\mathbf{r}, t) + (\mu_0 \boldsymbol{p} \times \mathbf{V}) \cdot \mathbf{H}(\mathbf{r}, t). \tag{8}$$

The last term of this Lagrangian contains the magnetic dipole-moment $\mathbf{m}(t) = \mu_0 \boldsymbol{p}(t) \times \mathbf{V}(t)$, which is the relativistic companion[5-8] of the moving electric point-dipole $\boldsymbol{p}(t)$.[†] The equations of motion may now be straightforwardly derived, using the standard technique,[4] as follows:

$$\partial \mathcal{L}/\partial \dot{x} = \left[ mV_x / \sqrt{1 - (V/c)^2} \right] + qA_x(\mathbf{r}, t) + \mu_0 p_z H_y - \mu_0 p_y H_z. \tag{9a}$$

$$\frac{d}{dt}(\partial \mathcal{L}/\partial \dot{x}) = \frac{d}{dt}\left[ mV_x / \sqrt{1 - (V/c)^2} \right] + q\left(\frac{\partial A_x}{\partial t} + \frac{\partial A_x}{\partial x}V_x + \frac{\partial A_x}{\partial y}V_y + \frac{\partial A_x}{\partial z}V_z\right)$$

$$+ \mu_0 \frac{\partial p_z}{\partial t} H_y + \mu_0 p_z \left(\frac{\partial H_y}{\partial t} + \frac{\partial H_y}{\partial x}V_x + \frac{\partial H_y}{\partial y}V_y + \frac{\partial H_y}{\partial z}V_z\right)$$

$$- \mu_0 \frac{\partial p_y}{\partial t} H_z - \mu_0 p_y \left(\frac{\partial H_z}{\partial t} + \frac{\partial H_z}{\partial x}V_x + \frac{\partial H_z}{\partial y}V_y + \frac{\partial H_z}{\partial z}V_z\right). \tag{9b}$$

$$\partial \mathcal{L}/\partial x = -q\frac{\partial \psi}{\partial x} + qV_x\frac{\partial A_x}{\partial x} + qV_y\frac{\partial A_y}{\partial x} + qV_z\frac{\partial A_z}{\partial x} + p_x\frac{\partial E_x}{\partial x} + p_y\frac{\partial E_y}{\partial x} + p_z\frac{\partial E_z}{\partial x}$$

$$+ \mu_0(p_y V_z - p_z V_y)\frac{\partial H_x}{\partial x} + \mu_0(p_z V_x - p_x V_z)\frac{\partial H_y}{\partial x} + \mu_0(p_x V_y - p_y V_x)\frac{\partial H_z}{\partial x}. \tag{9c}$$

Equating the last two identities, we find

$$\frac{dp_x}{dt} = q\left(-\frac{\partial \psi}{\partial x} - \frac{\partial A_x}{\partial t}\right) + qV_y\left(\frac{\partial A_y}{\partial x} - \frac{\partial A_x}{\partial y}\right) - qV_z\left(\frac{\partial A_x}{\partial z} - \frac{\partial A_z}{\partial x}\right)$$

$$- \mu_0 \frac{\partial p_z}{\partial t} H_y - \mu_0 p_z \left(\frac{\partial H_y}{\partial x}V_x + \frac{\partial H_y}{\partial y}V_y + \frac{\partial H_y}{\partial z}V_z\right)$$

$$+ \mu_0 \frac{\partial p_y}{\partial t} H_z + \mu_0 p_y \left(\frac{\partial H_z}{\partial x}V_x + \frac{\partial H_z}{\partial y}V_y + \frac{\partial H_z}{\partial z}V_z\right)$$

$$+ (\boldsymbol{p} \cdot \boldsymbol{\nabla})E_x + p_y\left(\frac{\partial E_y}{\partial x} - \frac{\partial E_x}{\partial y} + \underbrace{\frac{\partial \mu_0 H_z}{\partial t}}_{0}\right) - p_z\left(\frac{\partial E_x}{\partial z} - \frac{\partial E_z}{\partial x} + \underbrace{\frac{\partial \mu_0 H_y}{\partial t}}_{0}\right) \quad \leftarrow \boxed{\boldsymbol{\nabla} \times \mathbf{E} = -\partial \mathbf{B}/\partial t}$$

$$+ (\mathbf{m} \cdot \boldsymbol{\nabla})H_x + m_y\underbrace{\left(\frac{\partial H_y}{\partial x} - \frac{\partial H_x}{\partial y}\right)}_{\varepsilon_0 \partial E_z/\partial t} - m_z\underbrace{\left(\frac{\partial H_x}{\partial z} - \frac{\partial H_z}{\partial x}\right)}_{\varepsilon_0 \partial E_y/\partial t}. \quad \leftarrow \boxed{\boldsymbol{\nabla} \times \mathbf{H} = \varepsilon_0 \partial \mathbf{E}/\partial t} \tag{10}$$

---

[†]We use a definition of the magnetic induction in which $\mathbf{B}(\mathbf{r}, t) = \mu_0 \mathbf{H}(\mathbf{r}, t) + \mathbf{M}(\mathbf{r}, t)$, where $\mu_0$ is the permeability of free space. In this convention, the magnetization $\mathbf{M}$ has the dimensions of the $B$-field, in which case the magnetic dipole moment of a current loop in the $xy$-plane, having area $A$ and carrying an electric current $I$, is given by $\mathbf{m} = \mu_0 I A \, \hat{\mathbf{z}}$.

We note that the moving electric dipole represents the following polarization distribution:

$$\boldsymbol{P}(\boldsymbol{r},t) = \boldsymbol{p}(t)\delta[x - x(t)]\delta[y - y(t)]\delta[z - z(t)]. \tag{11}$$

The force-density exerted by the $H$-field on this polarization distribution is given by

$$[(\partial\boldsymbol{P}/\partial t) \times \mu_0 \boldsymbol{H}(\boldsymbol{r},t)]_x = \mu_0\{(\dot{p}_y H_z - \dot{p}_z H_y)\delta[x - x(t)]\delta[y - y(t)]\delta[z - z(t)]$$
$$-(p_y H_z - p_z H_y)V_x \delta'[x - x(t)]\delta[y - y(t)]\delta[z - z(t)]$$
$$-(p_y H_z - p_z H_y)V_y \delta[x - x(t)]\delta'[y - y(t)]\delta[z - z(t)]$$
$$-(p_y H_z - p_z H_y)V_z \delta[x - x(t)]\delta[y - y(t)]\delta'[z - z(t)]\}. \tag{12}$$

Integrating the above force-density over the entire $xyz$ space yields the force exerted by the $H$-field on the electric point-dipole, as follows:

$$\iiint_{\text{all space}} [(\partial\boldsymbol{P}/\partial t) \times \mu_0 \boldsymbol{H}(\boldsymbol{r},t)]_x \mathrm{d}x\mathrm{d}y\mathrm{d}z = \mu_0 \left( \frac{\partial p_y}{\partial t} H_z - \frac{\partial p_z}{\partial t} H_y + p_y V_x \frac{\partial H_z}{\partial x} - p_z V_x \frac{\partial H_y}{\partial x} \right.$$
$$\left. + p_y V_y \frac{\partial H_z}{\partial y} - p_z V_y \frac{\partial H_y}{\partial y} + p_y V_z \frac{\partial H_z}{\partial z} - p_z V_z \frac{\partial H_y}{\partial z} \right). \tag{13}$$

To simplify the notation, we shall write the expression in Eq.(13) as $[(\partial\boldsymbol{p}/\partial t) \times \mu_0 \boldsymbol{H}]_x$. The equation of motion, Eq.(10), may now be recast in the following form:

$$\frac{\mathrm{d}p_x}{\mathrm{d}t} = qE_x + q(\boldsymbol{V} \times \boldsymbol{B})_x + (\boldsymbol{p} \cdot \boldsymbol{\nabla})E_x + (\boldsymbol{m} \cdot \boldsymbol{\nabla})H_x + \left(\frac{\partial \boldsymbol{p}}{\partial t} \times \mu_0 \boldsymbol{H}\right)_x + \left(\boldsymbol{m} \times \varepsilon_0 \frac{\partial \boldsymbol{E}}{\partial t}\right)_x. \tag{14}$$

Similar equations are found for the components of the EM force along the $y$ and $z$ axes. Thus, the equation of motion of the point-particle in an external EM field turns out to be

$$\frac{\mathrm{d}\boldsymbol{p}}{\mathrm{d}t} = q(\boldsymbol{E} + \boldsymbol{V} \times \mu_0 \boldsymbol{H}) + (\boldsymbol{p} \cdot \boldsymbol{\nabla})\boldsymbol{E} + (\boldsymbol{m} \cdot \boldsymbol{\nabla})\boldsymbol{H} + \frac{\partial \boldsymbol{p}}{\partial t} \times \mu_0 \boldsymbol{H} - \frac{\partial \boldsymbol{m}}{\partial t} \times \varepsilon_0 \boldsymbol{E} + \frac{\mathrm{d}}{\mathrm{d}t}(\boldsymbol{m} \times \varepsilon_0 \boldsymbol{E}). \tag{15}[‡]$$

$\boxed{\boldsymbol{B} = \mu_0 \boldsymbol{H}}$

The expression on the right-hand side of Eq.(15) is the Lorentz force on the moving particle [i.e., point-charge $q$, plus electric point-dipole $\boldsymbol{p}(t)$, plus induced magnetic dipole $\boldsymbol{m}(t)$], which is deliberately written in the form of the Einstein-Laub force[18] plus the contribution of the hidden momentum, $\boldsymbol{m} \times \varepsilon_0 \boldsymbol{E}$.[24] Thus, the Lagrangian of Eq.(8) is seen to correspond to the Lorentz force of an external EM field acting on a moving electric dipole (together with its magnetic counterpart) within free space. To confirm this finding, Appendix C shows that the same force as in Eq.(15) is obtained by a direct integration of the Lorentz force density over the spatial distributions of the electric charge and current associated with the moving dipole.

When using the Lagrangian of Eq.(8), one should take care to amend the equation of motion by removing the hidden momentum contribution from Eq.(15).[25-32] It would have been desirable, of course, if a Lagrangian existed that yielded an equation of motion based solely on the Einstein-Laub force. However, in the absence of such a Lagrangian, one must always remember to apply the aforementioned correction. (To belabor the obvious from a slightly different point of view, Appendix E shows that a direct calculation of the Einstein-Laub force acting on a moving electric dipole results in the Lorentz force minus the hidden momentum contribution.)

---

[‡]The term $(\partial \boldsymbol{m}/\partial t) \times \varepsilon_0 \boldsymbol{E}$ should be treated similarly to $(\partial \boldsymbol{p}/\partial t) \times \mu_0 \boldsymbol{H}$; that is, in accordance with Eq.(13).

**4. The Hamiltonian of the moving electric dipole**. We derive the canonical momentum $P(r,t)$ of the point-particle from Eq.(9a), as follows:

$$P(r,t) = \frac{mV}{\sqrt{1-(V/c)^2}} + qA(r,t) - \wp \times \mu_0 H. \tag{16}$$

The above equation yields the following expressions for the velocity $V(t)$ and the relativistic factor $\sqrt{1-(V/c)^2}$:

$$\frac{mV}{\sqrt{1-(V/c)^2}} = P - qA + \wp \times \mu_0 H \quad \rightarrow \quad \frac{m^2 V^2}{1-(V/c)^2} = (P - qA + \wp \times \mu_0 H)^2$$

$$\rightarrow \quad \frac{1}{1-(V/c)^2} = 1 + \frac{(P-qA+\wp\times\mu_0 H)^2}{(mc)^2}. \tag{17}$$

The Hamiltonian of the point-particle is thus given by[4]

$$\mathcal{H}(r,P,t) = \sum_\ell P_\ell \dot{x}_\ell - \mathcal{L}(r,\dot{r},t)$$

$$= P \cdot V + mc^2\sqrt{1-(V/c)^2} + q\psi(r,t) - qV \cdot A(r,t) - \wp \cdot E(r,t) - \mu_0(\wp \times V) \cdot H(r,t)$$

$$= mc^2\sqrt{1-(V/c)^2} + (P - qA + \wp \times \mu_0 H) \cdot V + q\psi - \wp \cdot E$$

$$= mc^2\sqrt{1-(V/c)^2} + m^{-1}\sqrt{1-(V/c)^2}(P - qA + \wp \times \mu_0 H)^2 + q\psi - \wp \cdot E$$

$$= mc^2\sqrt{1-(V/c)^2}\left[1 + \frac{(P-qA+\wp\times\mu_0 H)^2}{(mc)^2}\right] + q\psi - \wp \cdot E$$

$$= mc^2\left\{1 + \frac{[P(r,t) - qA(r,t) + \wp(t)\times\mu_0 H(r,t)]^2}{(mc)^2}\right\}^{1/2} + q\psi(r,t) - \wp(t) \cdot E(r,t). \tag{18}$$

The non-relativistic version of the above Hamiltonian for $V \ll c$ is obtained by retaining only the first two terms in the Taylor series expansion of the square root, namely,

$$\mathcal{H}(r,P,t) \cong mc^2 + \frac{1}{2m}(P - qA + \wp \times \mu_0 H)^2 + q\psi(r,t) - \wp(t) \cdot E(r,t). \tag{19}$$

The non-relativistic Hamiltonian of the point-particle turns out to be the sum of its rest energy $mc^2$, kinetic energy $p^2/2m$, the (scalar) potential energy $q\psi(r,t)$ of the charge, and the potential energy $-\wp(t) \cdot E(r,t)$ of the electric dipole.

**5. Elongation and rotation of the electric point-dipole**. Let the *scalar* function $\wp(t)$ represent the magnitude of the dipole moment, its orientation within the $xyz$ coordinate system being described by the polar and azimuthal angles $\theta(t)$ and $\varphi(t)$. The remaining degree of rotational freedom, $\chi(t)$, does not have an associated potential energy within the EM field and, as such, it will not be considered any further. Any kinetic energy and angular momentum associated with the internal rotation $\dot{\chi}(t)$ around the dipole's axis will thus be constants of the motion.

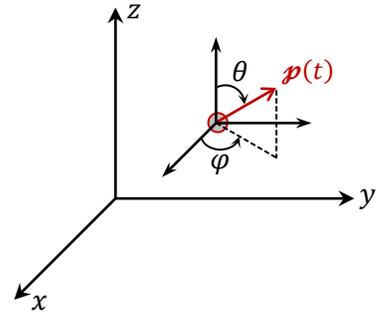

We invoke the simple mass-and-spring model[3,7] of a dipole oscillator having mass $m_e$, charge $e$, spring constant $\alpha$, and electric dipole moment $\wp(t)(\sin\theta\cos\varphi\,\hat{x} + \sin\theta\sin\varphi\,\hat{y} + \cos\theta\,\hat{z})$ to write down the Lagrangian of a charge-neutral electric dipole traveling within an external EM field, as follows:

$$\mathcal{L}(\mathbf{r},\dot{\mathbf{r}},\boldsymbol{p},\dot{\boldsymbol{p}},t) = -mc^2\sqrt{1-(V/c)^2} + \tfrac{1}{2}(m_e/e^2)\dot{p}^2 - \tfrac{1}{2}(\alpha/e^2)p^2 + \tfrac{1}{2}I_0\dot{\theta}^2 + \tfrac{1}{2}I_0\sin^2\theta\,\dot{\varphi}^2$$

$$+ p(t)(\sin\theta\cos\varphi\,\hat{\mathbf{x}} + \sin\theta\sin\varphi\,\hat{\mathbf{y}} + \cos\theta\,\hat{\mathbf{z}})\cdot \mathbf{E}(\mathbf{r},t)$$

$$+ \mu_0 p(t)[(\sin\theta\cos\varphi\,\hat{\mathbf{x}} + \sin\theta\sin\varphi\,\hat{\mathbf{y}} + \cos\theta\,\hat{\mathbf{z}})\times\mathbf{V}]\cdot\mathbf{H}(\mathbf{r},t). \qquad (20)$$

In the above equation, $m$ is the total mass of the dipole, while $m_e$ is the mass of its mobile (oscillating) part, whose electric charge is denoted by $e$. (By implication then, the massive part of the dipole has mass $m - m_e$ and charge $-e$.) The moment of inertia of the dipole for rotations involving a change of $\theta$ and/or $\varphi$ is denoted by $I_0$. (Imagining the dipole as a tumbling dumbbell clarifies the reason for setting both moments of inertia equal to $I_0$.) The equation of motion for the magnitude $p(t)$ of the dipole is thus found to be

$$\frac{d}{dt}\left(\frac{\partial \mathcal{L}}{\partial \dot{p}}\right) = \frac{\partial \mathcal{L}}{\partial p} \rightarrow (m_e/e^2)\ddot{p} = -(\alpha/e^2)p + (\sin\theta\cos\varphi\,\hat{\mathbf{x}} + \sin\theta\sin\varphi\,\hat{\mathbf{y}} + \cos\theta\,\hat{\mathbf{z}})\cdot\mathbf{E}(\mathbf{r},t)$$

$$+ \mu_0[(\sin\theta\cos\varphi\,\hat{\mathbf{x}} + \sin\theta\sin\varphi\,\hat{\mathbf{y}} + \cos\theta\,\hat{\mathbf{z}})\times\mathbf{V}]\cdot\mathbf{H}(\mathbf{r},t). \qquad (21)$$

This is just the equation of motion for the Lorentz oscillator model of an electric dipole,[3,7] which may be further streamlined [using the vector identity $(\mathbf{a}\times\mathbf{b})\cdot\mathbf{c} = \mathbf{a}\cdot(\mathbf{b}\times\mathbf{c})$], as follows:

$$\ddot{p} + (\alpha/m_e)p = (e^2/m_e)(\sin\theta\cos\varphi\,\hat{\mathbf{x}} + \sin\theta\sin\varphi\,\hat{\mathbf{y}} + \cos\theta\,\hat{\mathbf{z}})\cdot[\mathbf{E}(\mathbf{r},t) + \mathbf{V}\times\mu_0\mathbf{H}(\mathbf{r},t)]. \qquad (22)$$

Note that the resonance frequency appears in the above equation as $\omega_r = \sqrt{\alpha/m_e}$, as usual, whereas the damping coefficient (typically denoted by $\gamma$) is missing; the Lagrangian method cannot handle loss mechanisms which are generally rooted in dissipative processes akin to frictional forces. Next, we use the Lagrangian in Eq.(20) to derive the equation of motion for the polar angle $\theta$:

$$\frac{d}{dt}\left(\frac{\partial \mathcal{L}}{\partial \dot{\theta}}\right) = \frac{\partial \mathcal{L}}{\partial \theta} \rightarrow I_0\ddot{\theta} = I_0\sin\theta\cos\theta\,\dot{\varphi}^2 + p(t)(\cos\theta\cos\varphi\,\hat{\mathbf{x}} + \cos\theta\sin\varphi\,\hat{\mathbf{y}} - \sin\theta\,\hat{\mathbf{z}})\cdot\mathbf{E}(\mathbf{r},t)$$

$$+ \mu_0 p(t)[(\cos\theta\cos\varphi\,\hat{\mathbf{x}} + \cos\theta\sin\varphi\,\hat{\mathbf{y}} - \sin\theta\,\hat{\mathbf{z}})\times\mathbf{V}]\cdot\mathbf{H}(\mathbf{r},t). \qquad (23)$$

The first term on the right-hand side of Eq.(23) is the torque produced by the centrifugal force associated with the rotation of the frame of reference around the $z$-axis.[33] The motion of the particle in the $\varphi$-direction introduces a rotating frame, thus requiring the addition of the (fictitious) centrifugal torque to the EM torque along $\hat{\boldsymbol{\varphi}} = -\sin\varphi\,\hat{\mathbf{x}} + \cos\varphi\,\hat{\mathbf{y}}$ that acts on the moving particle. The second term, when written as $p(t)(E_x\cos\theta\cos\varphi + E_y\cos\theta\sin\varphi - E_z\sin\theta)$, is seen to be the projection of $\boldsymbol{p}(t)\times\mathbf{E}(\mathbf{r},t)$ on $\hat{\boldsymbol{\varphi}}$, precisely as expected from the expression of the EM torque acting on the point dipole; see the first term on the right-hand side of Eq.(D1) or Eq.(D4) of Appendix D. The last term in Eq.(23), which may be written as

$$\mu_0 p(t)\bigl[(H_xV_y - H_yV_x)\sin\theta + (H_xV_z - H_zV_x)\cos\theta\sin\varphi + (H_zV_y - H_yV_z)\cos\theta\cos\varphi\bigr], \qquad (24)$$

is the projection of $\boldsymbol{p}(t)\times(\mathbf{V}\times\mathbf{B})$ on $\hat{\boldsymbol{\varphi}}$, again in accordance with Eq.(D1), which is the expression of the EM torque acting at the location of the point-dipole. (See the discussion immediately following Eq.(D4) in Appendix D.)

Next, we use the Lagrangian of Eq.(20) to derive the equation of motion for the azimuthal angle $\varphi$:

$$\frac{d}{dt}\left(\frac{\partial \mathcal{L}}{\partial \dot{\varphi}}\right) = \frac{\partial \mathcal{L}}{\partial \varphi} \rightarrow I_0\sin^2\theta\,\ddot{\varphi} + 2I_0\sin\theta\cos\theta\,\dot{\theta}\dot{\varphi} = p(t)\sin\theta\,(-\sin\varphi\,\hat{\mathbf{x}} + \cos\varphi\,\hat{\mathbf{y}})\cdot\mathbf{E}(\mathbf{r},t)$$

$$+ \mu_0 p(t)\sin\theta\,[(-\sin\varphi\,\hat{\mathbf{x}} + \cos\varphi\,\hat{\mathbf{y}})\times\mathbf{V}]\cdot\mathbf{H}(\mathbf{r},t). \qquad (25)$$

On the left-hand side of this equation, the second term is due to the (fictitious) Coriolis force associated with the rotation of the frame around the z-axis.[33] On the right-hand-side, the first term is $(\boldsymbol{p} \times \boldsymbol{E}) \cdot \hat{\boldsymbol{z}}$, and the second term, which may be written as

$$\mu_0 \wp(t \sin\theta)[(H_x V_z - H_z V_x)\cos\varphi + (H_y V_z - H_z V_y)\sin\varphi], \tag{26}$$

is the projection of $\boldsymbol{p}(t) \times (\boldsymbol{V} \times \boldsymbol{B})$ on $\hat{\boldsymbol{z}}$, in agreement with Eq.(D1) of Appendix D. The total Lorentz torque acting on the moving electric dipole $\boldsymbol{p}(t)$ is thus seen to be $\boldsymbol{p}(t) \times (\boldsymbol{E} + \boldsymbol{V} \times \boldsymbol{B})$, where $\boldsymbol{V}(t) = \dot{\boldsymbol{r}}_p(t)$ is the instantaneous velocity of the particle, and the $\boldsymbol{E}$ and $\boldsymbol{B}$ fields are evaluated at the location of the particle at time $t$.

Finally, note that the hidden momentum $\boldsymbol{m} \times \varepsilon_0 \boldsymbol{E} = (\boldsymbol{p} \times \boldsymbol{V}) \times \boldsymbol{E}/c^2$ has an angular momentum counterpart that is given by $\boldsymbol{r}_p(t) \times (\boldsymbol{m} \times \varepsilon_0 \boldsymbol{E})$. The time-rate-of-change of this hidden angular momentum contributes a term $\boldsymbol{V} \times (\boldsymbol{m} \times \varepsilon_0 \boldsymbol{E}) = (\boldsymbol{E} \cdot \boldsymbol{V})(\boldsymbol{p} \times \boldsymbol{V})/c^2$ to the local torque acting on the particle, which must be removed from the Lorentz torque — in the same way that the hidden (linear) momentum's contribution had to be removed from Eq.(15).[22] The effective torque on the moving electric dipole is thus given by

$$\boldsymbol{T}(\boldsymbol{r}_p, t) = \boldsymbol{p}(t) \times (\boldsymbol{E} + \boldsymbol{V} \times \boldsymbol{B}) - (\boldsymbol{E} \cdot \boldsymbol{V})(\boldsymbol{p} \times \boldsymbol{V})/c^2. \tag{27}$$

This torque, which is the Lorentz torque minus the contribution from the hidden angular momentum, can be shown to be precisely what one gets from the Einstein-Laub torque equation.[24]

**6. Concluding remarks**. We have examined the equations of motion of a point-particle that carries an electric charge as well an electric dipole moment in the presence of an external EM field. It was shown that a simple Lagrangian that includes one term ($\boldsymbol{p} \cdot \boldsymbol{E}$) for the electric dipole and a second term ($\boldsymbol{m} \cdot \boldsymbol{H}$) for the relativistically-induced magnetic dipole of the particle, leads to an expression for the EM force that is precisely the expression that one obtains by a direct evaluation of the Lorentz force acting on the electric charge and current distributions of the moving particle. We pointed out that this expression of the Lorentz force inevitably contains a contribution from the hidden momentum that must subsequently be removed from the equations of motion. This is in contradistinction to the Einstein-Laub formulation[18] of the EM force acting on electric and magnetic dipoles, which automatically discounts such hidden momentum contributions.[22-32] Unfortunately, a corresponding Lagrangian for the Einstein-Laub formalism does not seem to exist and, as such, one must continue to be aware of the presence of the hidden momentum $\boldsymbol{m} \times \varepsilon_0 \boldsymbol{E}$ in theoretical studies that are based on the $\boldsymbol{p} \cdot \boldsymbol{E} + \boldsymbol{m} \cdot \boldsymbol{H}$ interaction Lagrangian.

We have pointed out (in Appendix B) that a straightforward evaluation of the Lagrangian of a moving electric dipole based solely on its electric charge and current distributions leads to the $\boldsymbol{p} \cdot \boldsymbol{E} + \boldsymbol{m} \cdot \boldsymbol{H}$ interaction Lagrangian in addition to a total time-derivative. In general, total time-derivatives that involve functions of position coordinates and time, can be removed from any Lagrangian without affecting the corresponding equations of motion. However, it is important to keep in mind that retaining a total time-derivative as part of the overall expression of the Lagrangian (much the same as a choice of gauge) could affect subsequent developments such as the form acquired by the canonical momentum or the expression obtained for the Hamiltonian of the system under consideration.[4]

The interaction Lagrangian $\boldsymbol{p} \cdot \boldsymbol{E} + \boldsymbol{m} \cdot \boldsymbol{H}$ also leads to equations for the rotational motion of the dipole pair, considering that the conjoined dipoles $\boldsymbol{p}(t)$ and $\boldsymbol{m}(t)$ are subject to a torque exerted by the external EM fields $\boldsymbol{E}(\boldsymbol{r},t)$ and $\boldsymbol{B}(\boldsymbol{r},t) = \mu_0 \boldsymbol{H}(\boldsymbol{r},t)$. In Sec.5, the formula $\boldsymbol{T} = \boldsymbol{p} \times (\boldsymbol{E} + \boldsymbol{V} \times \boldsymbol{B})$ was derived for the local and instantaneous EM torque acting on the particle moving with velocity $V(t)$ in otherwise free space. This is *not* the same torque as $p \times E + m \times H$ that one might naively have expected; the reasons for the discrepancy are discussed in Appendix D, where a direct calculation of the EM torque based on the Lorentz force law confirms the results obtained in Sec.5. We also argued that the term $(E \cdot V)(p \times V)/c^2$ must be subtracted from the Lorentz torque in order to eliminate a contribution by the hidden angular momentum. The effective torque acting on a moving electric dipole $p(t)$ together with its magnetic counterpart $m(t)$ is thus given by Eq.(27).

Extending the above results to a moving magnetic dipole (along with its relativistically-induced electric-dipole companion) runs into difficulties because, in this case, the function whose total time-derivative appears in the interaction Lagrangian is no longer a function of time and the particle coordinates only, but it also depends on the particle velocity $V(t)$. The issues associated with moving magnetic dipoles are discussed in Appendix F.

Throughout the paper, we have assumed that the EM fields $E$ and $B$ (and the corresponding potentials $\psi$ and $A$) are external, meaning that the self-fields of the point-particle (produced by its electric charge as well as its electric and magnetic dipoles) have been totally ignored. It is possible, of course, to include the self-fields in the analysis, but then the Lagrangian of the EM fields must be added to the sum of the particle Lagrangian and the interaction Lagrangian, as is routinely done in quantum electrodynamics studies.[1,4] Also sidestepped in our analyses are any discussion of the role played by the gauge, or by any gauge transformation, in arriving at the final results. This is because our scalar and vector potentials, $\psi$ and $A$, could be in any gauge without affecting the results obtained in this paper.

**Acknowledgement**. The author is grateful to Ewan M. Wright (College of Optical Sciences, University of Arizona), Arash Mafi (University of New Mexico), and Vladimir Hnizdo for several helpful discussions.

## Appendix A

### Electric charge and current densities of a moving electric dipole

A moving electric point-dipole has the polarization $\mathcal{P}(r,t) = p(t)\delta[r - r_p(t)]$, accompanied by the magnetization $\mathcal{M}(r,t) = \mu_0 \mathcal{P}(r,t) \times V(t)$.[5-8] The electric charge-density of this moving dipole pair is $\rho(r,t) = -\nabla \cdot \mathcal{P}(r,t)$, while its electric current-density is given by

$$J(r,t) = \partial \mathcal{P}/\partial t + \mu_0^{-1}\nabla \times \mathcal{M} = \dot{p}(t)\delta[r - r_p(t)] - (V \cdot \nabla)\mathcal{P} + \nabla \times (\mathcal{P} \times V)$$

$$\boxed{\nabla \times (a \times b) = a(\nabla \cdot b) - b(\nabla \cdot a) + (b \cdot \nabla)a - (a \cdot \nabla)b}$$

$$= \dot{p}(t)\delta[r - r_p(t)] - \cancel{(V \cdot \nabla)\mathcal{P}} - (\nabla \cdot \mathcal{P})V + \cancel{(V \cdot \nabla)\mathcal{P}}$$

$$= \dot{p}(t)\delta[r - r_p(t)] - (\nabla \cdot \mathcal{P})V(t). \tag{A1}$$

In this equation, the first term is due to the "elongation" and/or rotation of the dipole moment $p(t)$, whereas the second term corresponds to the motion of the electric dipole's pair of equal and opposite charges with velocity $V(t) = \dot{r}_p(t)$. The simple relation between the charge and current densities of the moving electric dipole given by Eq.(A1) should clarify the reason for the necessity of having a co-located magnetic dipole $m(t)$ accompany the electric dipole $p(t)$ at $r = r_p(t)$.

## Appendix B

### Deriving a Lagrangian for a moving electric dipole based on its charge and current densities

In the presence of an external EM field specified by the scalar potential $\psi(\mathbf{r},t)$ and the vector potential $\mathbf{A}(\mathbf{r},t)$, a Lagrangian for the point-dipole $\boldsymbol{p}(t)$ moving along an arbitrary trajectory $\mathbf{r}_p(t)$ while accompanied by a relativistically-induced magnetic dipole $\boldsymbol{m}(t) = \mu_0 \boldsymbol{p}(t) \times \dot{\mathbf{r}}_p(t)$, may be written down straightforwardly in terms of the electric charge density $\rho(\mathbf{r},t) = -\boldsymbol{\nabla} \cdot \boldsymbol{\mathcal{P}}(\mathbf{r},t)$ and the electric current density $\boldsymbol{J}(\mathbf{r},t) = \partial \boldsymbol{\mathcal{P}}/\partial t + \mu_0^{-1}\boldsymbol{\nabla} \times \boldsymbol{\mathcal{M}}$, as follows:

$$\mathcal{L}(\mathbf{r}_p, \dot{\mathbf{r}}_p, t) = \iiint_{\text{all space}} [(\boldsymbol{\nabla} \cdot \boldsymbol{\mathcal{P}})\psi(\mathbf{r},t) + (\partial \boldsymbol{\mathcal{P}}/\partial t + \mu_0^{-1}\boldsymbol{\nabla} \times \boldsymbol{\mathcal{M}}) \cdot \boldsymbol{A}(\mathbf{r},t)] \mathrm{d}x \mathrm{d}y \mathrm{d}z$$

$$\boxed{\boldsymbol{\nabla} \cdot (\psi \boldsymbol{\mathcal{P}}) = \psi \boldsymbol{\nabla} \cdot \boldsymbol{\mathcal{P}} + \boldsymbol{\mathcal{P}} \cdot \boldsymbol{\nabla} \psi} \quad \boxed{\boldsymbol{\nabla} \cdot (\boldsymbol{\mathcal{M}} \times \boldsymbol{A}) = \boldsymbol{A} \cdot (\boldsymbol{\nabla} \times \boldsymbol{\mathcal{M}}) - \boldsymbol{\mathcal{M}} \cdot (\boldsymbol{\nabla} \times \boldsymbol{A})}$$

$$= \int_{\text{all space}} [\boldsymbol{\nabla} \cdot (\psi \boldsymbol{\mathcal{P}} + \mu_0^{-1} \boldsymbol{\mathcal{M}} \times \boldsymbol{A}) - \boldsymbol{\mathcal{P}} \cdot \boldsymbol{\nabla}\psi - \boldsymbol{\mathcal{P}} \cdot (\partial \boldsymbol{A}/\partial t) + \partial(\boldsymbol{\mathcal{P}} \cdot \boldsymbol{A})/\partial t + \mu_0^{-1} \boldsymbol{\mathcal{M}} \cdot (\boldsymbol{\nabla} \times \boldsymbol{A})] \mathrm{d}v$$

$$= \iint_{\substack{\text{surface}\\\text{at }\infty}} (\psi \boldsymbol{\mathcal{P}} + \mu_0^{-1} \boldsymbol{\mathcal{M}} \times \boldsymbol{A}) \cdot \mathrm{d}\boldsymbol{s} + \int_{\text{all space}} (\boldsymbol{\mathcal{P}} \cdot \boldsymbol{E} + \mu_0^{-1} \boldsymbol{\mathcal{M}} \cdot \boldsymbol{B}) \mathrm{d}v + \frac{\mathrm{d}}{\mathrm{d}t}\int_{\text{all space}} (\boldsymbol{\mathcal{P}} \cdot \boldsymbol{A}) \mathrm{d}v$$

$$= \boldsymbol{p}(t) \cdot \boldsymbol{E}(\mathbf{r}_p, t) + \boldsymbol{m}(t) \cdot \boldsymbol{H}(\mathbf{r}_p, t) + \frac{\mathrm{d}}{\mathrm{d}t}\int_{\text{all space}} (\boldsymbol{\mathcal{P}} \cdot \boldsymbol{A}) \mathrm{d}v. \tag{B1}$$

The last term of the above Lagrangian, being a total time-derivative of a function of position coordinates and time, does not affect the equation of motion of the particle.[1] The remaining terms have been used in sections 3 and 5 to derive the relevant equations of motion.

## Appendix C

### Force of an external electromagnetic field on a moving electric point-dipole

The total force exerted by an EM field on a moving electric point-dipole (and its accompanying magnetic dipole) can be calculated based on the Lorentz law of force and a knowledge of the total electric charge and current distributions, which were described in Appendix A. We write

$$\boldsymbol{F} = \int_{\text{all space}} \{\dot{\boldsymbol{p}}(t)\delta[\mathbf{r} - \mathbf{r}_p(t)] \times \boldsymbol{B} - (\boldsymbol{\nabla} \cdot \boldsymbol{\mathcal{P}})(\boldsymbol{E} + \boldsymbol{V} \times \boldsymbol{B})\} \mathrm{d}v = \dot{\boldsymbol{p}}(t) \times \boldsymbol{B} + (\boldsymbol{p} \cdot \boldsymbol{\nabla})(\boldsymbol{E} + \boldsymbol{V} \times \boldsymbol{B})$$

$$\boxed{\text{Integration by parts}}$$

$$= (\boldsymbol{p} \cdot \boldsymbol{\nabla})\boldsymbol{E} + (\boldsymbol{p} \cdot \boldsymbol{\nabla})(\boldsymbol{V} \times \boldsymbol{B}) + \dot{\boldsymbol{p}}(t) \times \boldsymbol{B}$$

$$\boxed{\boldsymbol{\nabla}(\boldsymbol{a} \cdot \boldsymbol{b}) = (\boldsymbol{a} \cdot \boldsymbol{\nabla})\boldsymbol{b} + (\boldsymbol{b} \cdot \boldsymbol{\nabla})\boldsymbol{a} + \boldsymbol{a} \times (\boldsymbol{\nabla} \times \boldsymbol{b}) + \boldsymbol{b} \times (\boldsymbol{\nabla} \times \boldsymbol{a})}$$

$$= (\boldsymbol{p} \cdot \boldsymbol{\nabla})\boldsymbol{E} + \boldsymbol{\nabla}[\boldsymbol{p} \cdot (\boldsymbol{V} \times \boldsymbol{B})] - \boldsymbol{p} \times [\boldsymbol{\nabla} \times (\boldsymbol{V} \times \boldsymbol{B})] + \dot{\boldsymbol{p}}(t) \times \boldsymbol{B}$$

$$\boxed{\boldsymbol{a} \cdot (\boldsymbol{b} \times \boldsymbol{c}) = (\boldsymbol{a} \times \boldsymbol{b}) \cdot \boldsymbol{c}} \quad \boxed{\boldsymbol{\nabla} \times (\boldsymbol{a} \times \boldsymbol{b}) = \boldsymbol{a}(\boldsymbol{\nabla} \cdot \boldsymbol{b}) - \boldsymbol{b}(\boldsymbol{\nabla} \cdot \boldsymbol{a}) + (\boldsymbol{b} \cdot \boldsymbol{\nabla})\boldsymbol{a} - (\boldsymbol{a} \cdot \boldsymbol{\nabla})\boldsymbol{b}}$$

$$= (\boldsymbol{p} \cdot \boldsymbol{\nabla})\boldsymbol{E} + \boldsymbol{\nabla}[(\boldsymbol{p} \times \boldsymbol{V}) \cdot \boldsymbol{B}] - \boldsymbol{p} \times [\boldsymbol{V}(\boldsymbol{\nabla} \cdot \boldsymbol{B}) - (\boldsymbol{V} \cdot \boldsymbol{\nabla})\boldsymbol{B}] + \dot{\boldsymbol{p}}(t) \times \boldsymbol{B}$$

$$\boxed{\mu_0 \boldsymbol{p} \times \boldsymbol{V} = \boldsymbol{m}; \; \boldsymbol{B} = \mu_0 \boldsymbol{H};} \quad \boxed{\boldsymbol{\nabla} \cdot \boldsymbol{B} = 0}$$

$$= (\boldsymbol{p} \cdot \boldsymbol{\nabla})\boldsymbol{E} + [(\boldsymbol{m} \cdot \boldsymbol{\nabla})\boldsymbol{H} + \boldsymbol{m} \times (\boldsymbol{\nabla} \times \boldsymbol{H})] + [(\boldsymbol{V} \cdot \boldsymbol{\nabla})(\boldsymbol{p} \times \boldsymbol{B}) + \dot{\boldsymbol{p}}(t) \times \boldsymbol{B}]$$

$$\boxed{\boldsymbol{p} \times, \text{ being independent of } x, y, z, \text{ has been moved under the } \boldsymbol{\nabla} \text{ operator}}$$

$$= (\boldsymbol{p} \cdot \boldsymbol{\nabla})\boldsymbol{E} + (\boldsymbol{m} \cdot \boldsymbol{\nabla})\boldsymbol{H} + \boldsymbol{m} \times (\varepsilon_0 \partial \boldsymbol{E}/\partial t) + \int_{\text{all space}} (\partial \boldsymbol{\mathcal{P}}/\partial t) \times \mu_0 \boldsymbol{H}. \tag{C1}$$

$$\boxed{\boldsymbol{\nabla} \times \boldsymbol{H} = \varepsilon_0 \partial \boldsymbol{E}/\partial t} \quad \boxed{\boldsymbol{\mathcal{P}}(\mathbf{r},t) = \boldsymbol{p}(t)\delta[\mathbf{r} - \mathbf{r}_p(t)]}$$

For the moving electric dipole $\boldsymbol{p}(t)$ accompanied by the (relativistically-induced) magnetic dipole $\boldsymbol{m}(t)$, the preceding result is the same as that obtained in Sec.3, Eq.(15), which is based on the Lagrangian of Eq.(8).

## Appendix D

### Torque exerted by an electromagnetic field on a moving electric point-dipole

Invoking the total electric charge and current densities of a moving electric dipole (as described in Appendix A), one may compute the total torque experienced by the dipole in the following way:

$$\boldsymbol{T} = \int_{\text{all space}} \boldsymbol{r} \times \{\dot{\boldsymbol{p}}(t)\delta[\boldsymbol{r} - \boldsymbol{r}_p(t)] \times \boldsymbol{B}(\boldsymbol{r},t) - (\boldsymbol{\nabla} \cdot \boldsymbol{\mathcal{P}})(\boldsymbol{E} + \boldsymbol{V} \times \boldsymbol{B})\}\mathrm{d}v$$

$$= \boldsymbol{r}_p(t) \times [\dot{\boldsymbol{p}}(t) \times \boldsymbol{B}(\boldsymbol{r}_p,t)] - \int_{\text{all space}} (\boldsymbol{\nabla} \cdot \boldsymbol{\mathcal{P}})[\boldsymbol{r} \times (\boldsymbol{E} + \boldsymbol{V} \times \boldsymbol{B})]\mathrm{d}v$$

$$= \boldsymbol{r} \times [\dot{\boldsymbol{p}}(t) \times \boldsymbol{B}] + (\boldsymbol{p} \cdot \boldsymbol{\nabla})(\boldsymbol{r} \times \boldsymbol{E}) + (\boldsymbol{p} \cdot \boldsymbol{\nabla})[\boldsymbol{r} \times (\boldsymbol{V} \times \boldsymbol{B})] \quad \leftarrow \boxed{\partial \boldsymbol{r}/\partial x = \hat{\boldsymbol{x}};\ \partial \boldsymbol{r}/\partial y = \hat{\boldsymbol{y}};\ \partial \boldsymbol{r}/\partial z = \hat{\boldsymbol{z}}}$$

$$= \boldsymbol{p} \times \boldsymbol{E} + \boldsymbol{r} \times [(\boldsymbol{p} \cdot \boldsymbol{\nabla})\boldsymbol{E}] + \boldsymbol{p} \times (\boldsymbol{V} \times \boldsymbol{B}) + \boldsymbol{r} \times [(\boldsymbol{p} \cdot \boldsymbol{\nabla})(\boldsymbol{V} \times \boldsymbol{B})] + \boldsymbol{r} \times [\dot{\boldsymbol{p}}(t) \times \boldsymbol{B}]. \quad (D1)$$

The last two terms on the right-hand side of the above equation are related to $\partial_t \boldsymbol{\mathcal{P}}(\boldsymbol{r},t)$, as follows:

$$\int_{\text{all space}} \boldsymbol{r} \times (\partial_t \boldsymbol{\mathcal{P}} \times \boldsymbol{B})\mathrm{d}v = \boldsymbol{r}_p \times [\dot{\boldsymbol{p}}(t) \times \boldsymbol{B}] + (\boldsymbol{V} \cdot \boldsymbol{\nabla})[\boldsymbol{r} \times (\boldsymbol{p} \times \boldsymbol{B})]$$

$$= \boldsymbol{V} \times (\boldsymbol{p} \times \boldsymbol{B}) + \boldsymbol{r} \times [\boldsymbol{p} \times (\boldsymbol{V} \cdot \boldsymbol{\nabla})\boldsymbol{B}] + \boldsymbol{r} \times [\dot{\boldsymbol{p}}(t) \times \boldsymbol{B}]$$

$\boxed{\boldsymbol{p} \times, \text{ being independent of } x, y, z, \text{ has moved to the left of the } \boldsymbol{\nabla} \text{ operator}}$

$$= \boldsymbol{V} \times (\boldsymbol{p} \times \boldsymbol{B}) - \boldsymbol{r} \times \{\boldsymbol{p} \times [\boldsymbol{\nabla} \times (\boldsymbol{V} \times \boldsymbol{B}) - \boldsymbol{\nabla}(\boldsymbol{V} \cdot \boldsymbol{B})]\} + \boldsymbol{r} \times [\dot{\boldsymbol{p}}(t) \times \boldsymbol{B}]$$

$\boxed{\boldsymbol{\nabla}(\boldsymbol{a} \cdot \boldsymbol{b}) = (\boldsymbol{a} \cdot \boldsymbol{\nabla})\boldsymbol{b} + (\boldsymbol{b} \cdot \boldsymbol{\nabla})\boldsymbol{a} + \boldsymbol{a} \times (\boldsymbol{\nabla} \times \boldsymbol{b}) + \boldsymbol{b} \times (\boldsymbol{\nabla} \times \boldsymbol{a})}$

$$= \boldsymbol{V} \times (\boldsymbol{p} \times \boldsymbol{B}) - \boldsymbol{r} \times \{\boldsymbol{\nabla}[\boldsymbol{p} \cdot (\boldsymbol{V} \times \boldsymbol{B})] - (\boldsymbol{p} \cdot \boldsymbol{\nabla})(\boldsymbol{V} \times \boldsymbol{B})\} + \boldsymbol{r} \times [\dot{\boldsymbol{p}}(t) \times \boldsymbol{B}]$$

$\boxed{\boldsymbol{a} \cdot (\boldsymbol{b} \times \boldsymbol{c}) = (\boldsymbol{a} \times \boldsymbol{b}) \cdot \boldsymbol{c}} \leftarrow \boxed{\mu_0 \boldsymbol{p} \times \boldsymbol{V} = \boldsymbol{m};\ \boldsymbol{B} = \mu_0 \boldsymbol{H}}$

$$= \boldsymbol{V} \times (\boldsymbol{p} \times \boldsymbol{B}) - \boldsymbol{r} \times \boldsymbol{\nabla}(\boldsymbol{m} \cdot \boldsymbol{H}) + \boldsymbol{r} \times [(\boldsymbol{p} \cdot \boldsymbol{\nabla})(\boldsymbol{V} \times \boldsymbol{B})] + \boldsymbol{r} \times [\dot{\boldsymbol{p}}(t) \times \boldsymbol{B}]$$

$$= \boldsymbol{V} \times (\boldsymbol{p} \times \boldsymbol{B}) - \boldsymbol{r} \times [(\boldsymbol{m} \cdot \boldsymbol{\nabla})\boldsymbol{H} + \boldsymbol{m} \times \varepsilon_0 \partial_t \boldsymbol{E}] + \boldsymbol{r} \times [(\boldsymbol{p} \cdot \boldsymbol{\nabla})(\boldsymbol{V} \times \boldsymbol{B})] + \boldsymbol{r} \times [\dot{\boldsymbol{p}}(t) \times \boldsymbol{B}]. \quad (D2)$$

$\boxed{\boldsymbol{\nabla} \times \boldsymbol{H} = \varepsilon_0 \partial \boldsymbol{E}/\partial t}$

Substitution into Eq.(D1) now yields

$$\boldsymbol{T} = \boldsymbol{p} \times \boldsymbol{E} + \boldsymbol{r} \times [(\boldsymbol{p} \cdot \boldsymbol{\nabla})\boldsymbol{E}] + \boldsymbol{p} \times (\boldsymbol{V} \times \boldsymbol{B}) - \boldsymbol{V} \times (\boldsymbol{p} \times \boldsymbol{B}) + \boldsymbol{r} \times [(\boldsymbol{m} \cdot \boldsymbol{\nabla})\boldsymbol{H} + \boldsymbol{m} \times \varepsilon_0 \partial_t \boldsymbol{E}]$$

$$+ \int_{\text{all space}} \boldsymbol{r} \times (\partial_t \boldsymbol{\mathcal{P}} \times \boldsymbol{B})\mathrm{d}v. \quad (D3)$$

Noting that $\boxed{\boldsymbol{a} \times (\boldsymbol{b} \times \boldsymbol{c}) = (\boldsymbol{a} \cdot \boldsymbol{c})\boldsymbol{b} - (\boldsymbol{a} \cdot \boldsymbol{b})\boldsymbol{c}}$

$$\boldsymbol{p} \times (\boldsymbol{V} \times \boldsymbol{B}) - \boldsymbol{V} \times (\boldsymbol{p} \times \boldsymbol{B}) = (\boldsymbol{p} \cdot \boldsymbol{B})\boldsymbol{V} - (\boldsymbol{p} \cdot \boldsymbol{V})\boldsymbol{B} - (\boldsymbol{V} \cdot \boldsymbol{B})\boldsymbol{p} + (\boldsymbol{V} \cdot \boldsymbol{p})\boldsymbol{B} = (\boldsymbol{p} \times \boldsymbol{V}) \times \boldsymbol{B} = \boldsymbol{m} \times \boldsymbol{H},$$

we finally arrive at

$$\boldsymbol{T} = \boldsymbol{p} \times \boldsymbol{E} + \boldsymbol{m} \times \boldsymbol{H} + \boldsymbol{r} \times [(\boldsymbol{p} \cdot \boldsymbol{\nabla})\boldsymbol{E} + (\boldsymbol{m} \cdot \boldsymbol{\nabla})\boldsymbol{H}] + \int_{\text{all space}} \boldsymbol{r} \times (\partial_t \boldsymbol{\mathcal{P}} \times \boldsymbol{B})\mathrm{d}v + \boldsymbol{r} \times [\boldsymbol{m} \times \varepsilon_0 \partial_t \boldsymbol{E}]. \quad (D4)$$

Notable are the similarities and differences between the force equation, Eq.(C1), and the corresponding torque equation, Eq.(D4). Note also that, at the location of the particle (i.e., at $\boldsymbol{r} = 0$), the torque is *not* given by $\boldsymbol{T}(\boldsymbol{r} = 0, t) = \boldsymbol{p} \times \boldsymbol{E} + \boldsymbol{m} \times \boldsymbol{H}$, as might erroneously be inferred from Eq.(D4), but rather by $\boldsymbol{T}(\boldsymbol{r} = 0, t) = \boldsymbol{p} \times (\boldsymbol{E} + \boldsymbol{V} \times \boldsymbol{B})$, in accordance with Eq.(D1).

## Appendix E

### Electromagnetic force acting on a moving electric dipole in the Einstein-Laub formulation

An electric point-dipole $\boldsymbol{p}(t)$ moves along the trajectory $\boldsymbol{r}_p(t) = x_p(t)\hat{\boldsymbol{x}} + y_p(t)\hat{\boldsymbol{y}} + z_p(t)\hat{\boldsymbol{z}}$ within an external electromagnetic (EM) field, i.e., one that excludes its own field (or self-field). The accompanying (relativistically-induced) magnetic dipole moment is $\boldsymbol{m}(t) = \mu_0 \boldsymbol{p}(t) \times \boldsymbol{V}(t)$, where $\boldsymbol{V}(t) = \dot{x}_p(t)\hat{\boldsymbol{x}} + \dot{y}_p(t)\hat{\boldsymbol{y}} + \dot{z}_p(t)\hat{\boldsymbol{z}}$. The highly-localized polarization and magnetization distributions are thus given by

$$\boldsymbol{\mathcal{P}}(\boldsymbol{r},t) = \boldsymbol{p}(t)\delta[x - x_p(t)]\delta[y - y_p(t)]\delta[z - z_p(t)]. \tag{E1}$$

$$\boldsymbol{\mathcal{M}}(\boldsymbol{r},t) = \boldsymbol{m}(t)\delta[x - x_p(t)]\delta[y - y_p(t)]\delta[z - z_p(t)]. \tag{E2}$$

We compute the EM force acting on the dipole pair in accordance with the Einstein-Laub force density, which, integrated over the shared volume of the dipoles, yields

$$\boldsymbol{F}_{\mathrm{EL}}(t) = \iiint_{-\infty}^{\infty}[(\boldsymbol{\mathcal{P}} \cdot \boldsymbol{\nabla})\boldsymbol{E} + \partial_t \boldsymbol{\mathcal{P}} \times \mu_0 \boldsymbol{H} + (\boldsymbol{\mathcal{M}} \cdot \boldsymbol{\nabla})\boldsymbol{H} - \partial_t \boldsymbol{\mathcal{M}} \times \varepsilon_0 \boldsymbol{E}] \mathrm{d}v. \tag{E3}$$

The individual terms appearing in the above equation may now be evaluated, as follows:

$$\int_{-\infty}^{\infty}(\boldsymbol{\mathcal{P}} \cdot \boldsymbol{\nabla})\boldsymbol{E}\,\mathrm{d}v = [\boldsymbol{p}(t) \cdot \boldsymbol{\nabla}]\boldsymbol{E}(\boldsymbol{r},t)|_{\boldsymbol{r}=\boldsymbol{r}_p(t)}. \tag{E4}$$

$$\int_{-\infty}^{\infty} \partial_t \boldsymbol{\mathcal{P}} \times \mu_0 \boldsymbol{H}\,\mathrm{d}v = \dot{\boldsymbol{p}}(t) \times \mu_0 \boldsymbol{H}(\boldsymbol{r},t)|_{\boldsymbol{r}=\boldsymbol{r}_p(t)} + \mu_0 \boldsymbol{p}(t) \times (\boldsymbol{V} \cdot \boldsymbol{\nabla})\boldsymbol{H}(\boldsymbol{r},t)|_{\boldsymbol{r}=\boldsymbol{r}_p(t)}. \tag{E5}$$

$$\int_{-\infty}^{\infty}(\boldsymbol{\mathcal{M}} \cdot \boldsymbol{\nabla})\boldsymbol{H}\,\mathrm{d}v = \{[\mu_0 \boldsymbol{p}(t) \times \boldsymbol{V}] \cdot \boldsymbol{\nabla}\}\boldsymbol{H}(\boldsymbol{r},t)|_{\boldsymbol{r}=\boldsymbol{r}_p(t)} \quad \boxed{\boldsymbol{\nabla}(\boldsymbol{a}\cdot\boldsymbol{b}) = (\boldsymbol{a}\cdot\boldsymbol{\nabla})\boldsymbol{b} + (\boldsymbol{b}\cdot\boldsymbol{\nabla})\boldsymbol{a} + \boldsymbol{a}\times(\boldsymbol{\nabla}\times\boldsymbol{b}) + \boldsymbol{b}\times(\boldsymbol{\nabla}\times\boldsymbol{a})}$$

$$= \mu_0 \boldsymbol{\nabla}[(\boldsymbol{p} \times \boldsymbol{V}) \cdot \boldsymbol{H}] - (\mu_0 \boldsymbol{p} \times \boldsymbol{V}) \times (\boldsymbol{\nabla} \times \boldsymbol{H})$$

$$\boxed{(\boldsymbol{a}\times\boldsymbol{b})\cdot\boldsymbol{c} = (\boldsymbol{b}\times\boldsymbol{c})\cdot\boldsymbol{a}} \to = \mu_0 \boldsymbol{\nabla}[(\boldsymbol{V} \times \boldsymbol{H}) \cdot \boldsymbol{p}] - \boldsymbol{m} \times \partial_t(\varepsilon_0 \boldsymbol{E}) \quad \boxed{\boldsymbol{D} = \varepsilon_0\boldsymbol{E} + \boldsymbol{P} = \varepsilon_0\boldsymbol{E} \text{ because self-field is excluded}}$$

$$= \mu_0(\boldsymbol{p} \cdot \boldsymbol{\nabla})(\boldsymbol{V} \times \boldsymbol{H}) + \mu_0 \boldsymbol{p} \times [\boldsymbol{\nabla} \times (\boldsymbol{V} \times \boldsymbol{H})] - \boldsymbol{m} \times \partial_t(\varepsilon_0 \boldsymbol{E})$$

$$\boxed{\boldsymbol{\nabla}\times(\boldsymbol{a}\times\boldsymbol{b}) = \boldsymbol{a}(\boldsymbol{\nabla}\cdot\boldsymbol{b}) - \boldsymbol{b}(\boldsymbol{\nabla}\cdot\boldsymbol{a}) + (\boldsymbol{b}\cdot\boldsymbol{\nabla})\boldsymbol{a} - (\boldsymbol{a}\cdot\boldsymbol{\nabla})\boldsymbol{b}}$$

$$= (\boldsymbol{p} \cdot \boldsymbol{\nabla})(\boldsymbol{V} \times \mu_0 \boldsymbol{H}) + \mu_0 \boldsymbol{p} \times [\boldsymbol{V}(\boldsymbol{\nabla} \cdot \boldsymbol{H}) - (\boldsymbol{V} \cdot \boldsymbol{\nabla})\boldsymbol{H}] - \boldsymbol{m} \times \partial_t(\varepsilon_0 \boldsymbol{E}). \tag{E6}$$

$$\boxed{\mu_0 \boldsymbol{\nabla}\cdot\boldsymbol{H} = \boldsymbol{\nabla}\cdot(\mu_0\boldsymbol{H} + \boldsymbol{M}) = \boldsymbol{\nabla}\cdot\boldsymbol{B} \text{ because self-field is excluded}}$$

$$\int_{-\infty}^{\infty}(\partial_t \boldsymbol{\mathcal{M}} \times \varepsilon_0 \boldsymbol{E})\,\mathrm{d}v = \dot{\boldsymbol{m}}(t) \times \varepsilon_0 \boldsymbol{E}(\boldsymbol{r},t)|_{\boldsymbol{r}=\boldsymbol{r}_p(t)} + \varepsilon_0 \boldsymbol{m}(t) \times (\boldsymbol{V} \cdot \boldsymbol{\nabla})\boldsymbol{E}(\boldsymbol{r},t)|_{\boldsymbol{r}=\boldsymbol{r}_p(t)}. \tag{E7}$$

Combining the above expressions, we find $\boxed{\boldsymbol{B} = \mu_0 \boldsymbol{H}}$

$$\boldsymbol{F}_{\mathrm{EL}}(t) = [\boldsymbol{p}(t) \cdot \boldsymbol{\nabla}](\boldsymbol{E} + \boldsymbol{V} \times \boldsymbol{B}) + \dot{\boldsymbol{p}}(t) \times \boldsymbol{B} - \frac{\mathrm{d}}{\mathrm{d}t}[\boldsymbol{m}(t) \times \varepsilon_0 \boldsymbol{E}]. \tag{E8}$$

The final expression in Eq.(E8) is the Lorentz force minus the hidden momentum contribution.

## Appendix F

### Electromagnetic force acting on a moving magnetic dipole in an external EM field

Let the magnetic point-dipole $\boldsymbol{m}(t)$ move along the trajectory $\boldsymbol{r}_p(t) = x_p(t)\hat{\boldsymbol{x}} + y_p(t)\hat{\boldsymbol{y}} + z_p(t)\hat{\boldsymbol{z}}$ under the influence of the scalar and vector potentials $\psi(\boldsymbol{r},t)$ and $\boldsymbol{A}(\boldsymbol{r},t)$ in an otherwise free space. The potentials are produced by external sources, which means that the self-field of the moving dipole is being ignored here. The relativistically-induced electric dipole moment accompanying the moving magnetic dipole is known to be $\boldsymbol{p}(t) = \varepsilon_0 \boldsymbol{V}(t) \times \boldsymbol{m}(t)$, where $\boldsymbol{V}(t) = \dot{\boldsymbol{r}}_p(t)$ is the point-particle's instantaneous velocity. The electric charge and current densities associated with the co-moving dipole pair are given by

$$\rho(\boldsymbol{r},t) = -\boldsymbol{\nabla} \cdot \boldsymbol{\mathcal{P}}(\boldsymbol{r},t) = -\boldsymbol{\nabla} \cdot \{\varepsilon_0 \boldsymbol{V} \times \underbrace{\boldsymbol{m}\delta[\boldsymbol{r} - \boldsymbol{r}_p(t)]}_{\boldsymbol{\mathcal{M}}(\boldsymbol{r},t)}\} = \varepsilon_0 \boldsymbol{V} \cdot (\boldsymbol{\nabla} \times \boldsymbol{\mathcal{M}}). \quad \text{(F1)}$$

$$\boxed{\boldsymbol{\nabla} \cdot (\boldsymbol{a} \times \boldsymbol{b}) = \boldsymbol{b} \cdot (\boldsymbol{\nabla} \times \boldsymbol{a}) - \boldsymbol{a} \cdot (\boldsymbol{\nabla} \times \boldsymbol{b})}$$

$$\boldsymbol{J}(\boldsymbol{r},t) = \mu_0^{-1} \boldsymbol{\nabla} \times \boldsymbol{\mathcal{M}} + \frac{\partial}{\partial t}\underbrace{(\varepsilon_0 \boldsymbol{V} \times \boldsymbol{\mathcal{M}})}_{\boldsymbol{\mathcal{P}}(\boldsymbol{r},t)}$$

$$= \mu_0^{-1} \boldsymbol{\nabla} \times \boldsymbol{\mathcal{M}} + \varepsilon_0 \dot{\boldsymbol{V}} \times \boldsymbol{\mathcal{M}} + (\varepsilon_0 \boldsymbol{V} \times \dot{\boldsymbol{m}})\delta[\boldsymbol{r} - \boldsymbol{r}_p(t)] - \varepsilon_0 \boldsymbol{V} \times [(\boldsymbol{V} \cdot \boldsymbol{\nabla})\boldsymbol{\mathcal{M}}]. \quad \text{(F2)}$$

The charge-current continuity equation, $\boldsymbol{\nabla} \cdot \boldsymbol{J} + \partial \rho / \partial t = 0$, may be straightforwardly verified from Eqs.(F1) and (F2). The following identity will be needed in the derivations that follow:

$$\frac{\mathrm{d}}{\mathrm{d}t} \int_{\text{all space}} (\boldsymbol{V} \times \boldsymbol{\mathcal{M}}) \cdot \boldsymbol{A}(\boldsymbol{r},t)\mathrm{d}v = (\dot{\boldsymbol{V}} \times \boldsymbol{m}) \cdot \boldsymbol{A}(\boldsymbol{r}_p,t) + (\boldsymbol{V} \times \dot{\boldsymbol{m}}) \cdot \boldsymbol{A}(\boldsymbol{r}_p,t) + (\boldsymbol{V} \times \boldsymbol{m}) \cdot \partial_t \boldsymbol{A}(\boldsymbol{r}_p,t)$$

$$- \int_{\text{all space}} \{\boldsymbol{V} \times [(\boldsymbol{V} \cdot \boldsymbol{\nabla})\boldsymbol{\mathcal{M}}]\} \cdot \boldsymbol{A}(\boldsymbol{r},t)\mathrm{d}v. \quad \text{(F3)}$$

In the presence of the scalar and vector potentials $\psi$ and $\boldsymbol{A}$, the standard interaction Lagrangian for the charge and current densities of Eqs.(F1) and (F2), may be written as

$$\mathcal{L}(\boldsymbol{r}_p, \dot{\boldsymbol{r}}_p, t) = -\int_{\text{all space}} [\varepsilon_0 \boldsymbol{V} \cdot (\boldsymbol{\nabla} \times \boldsymbol{\mathcal{M}})]\psi(\boldsymbol{r},t)\mathrm{d}v$$

$$+ \int_{\text{all space}} \{\mu_0^{-1}(\boldsymbol{\nabla} \times \boldsymbol{\mathcal{M}}) + \varepsilon_0(\dot{\boldsymbol{V}} \times \boldsymbol{\mathcal{M}}) + (\varepsilon_0 \boldsymbol{V} \times \dot{\boldsymbol{m}})\delta[\boldsymbol{r} - \boldsymbol{r}_p(t)] - \varepsilon_0\{\boldsymbol{V} \times [(\boldsymbol{V} \cdot \boldsymbol{\nabla})\boldsymbol{\mathcal{M}}]\}\} \cdot \boldsymbol{A}(\boldsymbol{r},t)\mathrm{d}v$$

$$\boxed{\boldsymbol{\nabla} \times \boldsymbol{A} = \boldsymbol{B} = \mu_0 \boldsymbol{H}}$$

$$= \varepsilon_0 \int \psi \boldsymbol{\nabla} \cdot (\boldsymbol{V} \times \boldsymbol{\mathcal{M}})\mathrm{d}v + \mu_0^{-1} \int \boldsymbol{\nabla} \cdot (\boldsymbol{\mathcal{M}} \times \boldsymbol{A})\mathrm{d}v \xrightarrow{0} + \mu_0^{-1} \int \boldsymbol{\mathcal{M}} \cdot (\boldsymbol{\nabla} \times \boldsymbol{A})\mathrm{d}v$$

$$+ \varepsilon_0(\dot{\boldsymbol{V}} \times \boldsymbol{m} + \boldsymbol{V} \times \dot{\boldsymbol{m}}) \cdot \boldsymbol{A}(\boldsymbol{r}_p,t) - \varepsilon_0 \int \{\boldsymbol{V} \times [(\boldsymbol{V} \cdot \boldsymbol{\nabla})\boldsymbol{\mathcal{M}}]\} \cdot \boldsymbol{A}(\boldsymbol{r},t)\mathrm{d}v$$

$$= \varepsilon_0 \int \boldsymbol{\nabla} \cdot [\psi(\boldsymbol{V} \times \boldsymbol{\mathcal{M}})]\mathrm{d}v \xrightarrow{0} - \varepsilon_0(\boldsymbol{V} \times \boldsymbol{m}) \cdot \boldsymbol{\nabla}\psi + \boldsymbol{m} \cdot \boldsymbol{H} + \varepsilon_0(\dot{\boldsymbol{V}} \times \boldsymbol{m} + \boldsymbol{V} \times \dot{\boldsymbol{m}}) \cdot \boldsymbol{A}(\boldsymbol{r}_p,t)$$

$$-\varepsilon_0 \underbrace{\int_{\text{all space}} \{\boldsymbol{V} \times [(\boldsymbol{V} \cdot \boldsymbol{\nabla})\boldsymbol{\mathcal{M}}]\} \cdot \boldsymbol{A}(\boldsymbol{r},t)\mathrm{d}v}_{\boldsymbol{E}(\boldsymbol{r}_p,t)} \leftarrow \text{Substitute for this term from Eq.(F3)}$$

$$= \varepsilon_0(\boldsymbol{V} \times \boldsymbol{m}) \cdot \overbrace{(-\boldsymbol{\nabla}\psi - \partial_t \boldsymbol{A})}^{\boldsymbol{E}(\boldsymbol{r}_p,t)} + \boldsymbol{m} \cdot \boldsymbol{H} + \frac{\mathrm{d}}{\mathrm{d}t}\int_{\text{all space}}(\varepsilon_0 \boldsymbol{V} \times \boldsymbol{\mathcal{M}}) \cdot \boldsymbol{A}(\boldsymbol{r},t)\mathrm{d}v$$

$$= \boldsymbol{p}(t) \cdot \boldsymbol{E}(\boldsymbol{r}_p,t) + \boldsymbol{m}(t) \cdot \boldsymbol{H}(\boldsymbol{r}_p,t) + \frac{\mathrm{d}}{\mathrm{d}t}\int_{\text{all space}}(\varepsilon_0 \boldsymbol{V} \times \boldsymbol{\mathcal{M}}) \cdot \boldsymbol{A}(\boldsymbol{r},t)\mathrm{d}v. \quad \text{(F4)}$$

Unfortunately, the function whose total time-derivative appears in Eq.(F4) depends not just on the particle coordinates $\boldsymbol{r}_p$, but also on the velocity $\dot{\boldsymbol{r}}_p = \boldsymbol{V}(t)$. Consequently, the simple expression $\boldsymbol{p} \cdot \boldsymbol{E} + \boldsymbol{m} \cdot \boldsymbol{H}$ cannot serve as an acceptable Lagrangian for a magnetic point-dipole moving within an external EM field.[4] This is readily verified by trying to determine the equation of motion from the reduced Lagrangian, and finding that the expression of the force acting on the moving pair of dipoles does not agree with the expected force as derived from the Lorentz law.

---

**Digression**. We show that the $\boldsymbol{p} \cdot \boldsymbol{E} + \boldsymbol{m} \cdot \boldsymbol{H}$ interaction Lagrangian does *not* yield the expected Lorentz force acting on the moving magnetic point-dipole $\boldsymbol{m}(t)$ accompanied by the relativistically-induced electric point-dipole $\boldsymbol{p}(t)$. The reduced Lagrangian and its derivatives are

$$\mathcal{L}(\boldsymbol{r}_p, \dot{\boldsymbol{r}}_p, t) = -mc^2\sqrt{1 - (V/c)^2} + \boldsymbol{m} \cdot \boldsymbol{H} + (\varepsilon_0 \boldsymbol{V} \times \boldsymbol{m}) \cdot \boldsymbol{E}$$

$$= -mc^2\sqrt{1 - (\boldsymbol{V} \cdot \boldsymbol{V}/c^2)} + \boldsymbol{m} \cdot \boldsymbol{H} + \varepsilon_0(\boldsymbol{m} \times \boldsymbol{E}) \cdot \boldsymbol{V}. \quad \text{(F5)}$$

$$\frac{d}{dt}\left(\frac{\partial \mathcal{L}}{\partial \dot{r}}\right) = \frac{d}{dt}\left(\frac{mV}{\sqrt{1-(V/c)^2}}\right) + \varepsilon_0 \frac{d}{dt}(\boldsymbol{m} \times \boldsymbol{E})$$

$$= \dot{\boldsymbol{p}} + \varepsilon_0(\dot{\boldsymbol{m}} \times \boldsymbol{E}) + \varepsilon_0 \boldsymbol{m} \times \partial_t \boldsymbol{E} + \varepsilon_0 \boldsymbol{m} \times (\boldsymbol{V} \cdot \boldsymbol{\nabla})\boldsymbol{E}. \tag{F6}$$

$$\frac{\partial \mathcal{L}}{\partial \boldsymbol{r}} = \boldsymbol{\nabla}\mathcal{L} = (\boldsymbol{m} \cdot \boldsymbol{\nabla})\boldsymbol{H} + \boldsymbol{m} \times (\boldsymbol{\nabla} \times \boldsymbol{H}) + \varepsilon_0 (\boldsymbol{V} \cdot \boldsymbol{\nabla})(\boldsymbol{m} \times \boldsymbol{E}) + \varepsilon_0 \boldsymbol{V} \times [\boldsymbol{\nabla} \times (\boldsymbol{m} \times \boldsymbol{E})]. \tag{F7}$$

Equating Eq.(F6) with Eq.(F7) yields the point-particle's equation of motion, as follows:

$$\dot{\boldsymbol{p}} = (\boldsymbol{m} \cdot \boldsymbol{\nabla})\boldsymbol{H} + \cancel{\boldsymbol{m} \times (\boldsymbol{\nabla} \times \boldsymbol{H})} + \varepsilon_0 \cancel{(\boldsymbol{V} \cdot \boldsymbol{\nabla})(\boldsymbol{m} \times \boldsymbol{E})} + \varepsilon_0 \boldsymbol{V} \times [\boldsymbol{\nabla} \times (\boldsymbol{m} \times \boldsymbol{E})]$$

[$\boldsymbol{\nabla} \times (\boldsymbol{a} \times \boldsymbol{b}) = (\boldsymbol{\nabla} \cdot \boldsymbol{b})\boldsymbol{a} - (\boldsymbol{\nabla} \cdot \boldsymbol{a})\boldsymbol{b} + (\boldsymbol{b} \cdot \boldsymbol{\nabla})\boldsymbol{a} - (\boldsymbol{a} \cdot \boldsymbol{\nabla})\boldsymbol{b}$]

$$-\varepsilon_0(\dot{\boldsymbol{m}} \times \boldsymbol{E}) - \cancel{\varepsilon_0 \boldsymbol{m} \times \partial_t \boldsymbol{E}} - \cancel{\varepsilon_0 \boldsymbol{m} \times (\boldsymbol{V} \cdot \boldsymbol{\nabla})\boldsymbol{E}}$$

[$\boldsymbol{m} \times$, being independent of $x, y, z$, has moved to the left of the $\boldsymbol{\nabla}$ operator.]

$$= (\boldsymbol{m} \cdot \boldsymbol{\nabla})\boldsymbol{H} - \varepsilon_0(\dot{\boldsymbol{m}} \times \boldsymbol{E}) - \varepsilon_0 \boldsymbol{V} \times [(\boldsymbol{\nabla} \cdot \boldsymbol{E})\boldsymbol{m}^{\nearrow 0} + (\boldsymbol{m} \cdot \boldsymbol{\nabla})\boldsymbol{E}]. \tag{F8}§$$

The right-hand side of Eq.(F8) represents neither the Lorentz force nor the Einstein-Laub force. For comparison, the various terms in the expression of the Lorentz force are listed below. (Subtracting the time derivative of $\int (\boldsymbol{\mathcal{M}} \times \varepsilon_0 \boldsymbol{E})dv$ from the Lorentz force would yield the Einstein-Laub force.)

$$\int_{\text{all space}} \rho(\boldsymbol{r},t)\boldsymbol{E}(\boldsymbol{r},t)dv = \int_{-\infty}^{\infty}(-\boldsymbol{\nabla} \cdot \boldsymbol{\mathcal{P}})\boldsymbol{E}dv \underset{\text{Integration by parts}}{=} \int_{-\infty}^{\infty}(\boldsymbol{\mathcal{P}} \cdot \boldsymbol{\nabla})\boldsymbol{E}dv = (\boldsymbol{p} \cdot \boldsymbol{\nabla})\boldsymbol{E} = \varepsilon_0[(\boldsymbol{V} \times \boldsymbol{m}) \cdot \boldsymbol{\nabla}]\boldsymbol{E}. \tag{F9}$$

$$\int_{\text{all space}} \underbrace{\frac{\partial}{\partial t}(\varepsilon_0 \boldsymbol{V} \times \boldsymbol{m})\delta[\boldsymbol{r} - \boldsymbol{r}_p(t)]}_{\boldsymbol{\mathcal{P}}(\boldsymbol{r},t)} \times \boldsymbol{B}dv = \varepsilon_0[(\dot{\boldsymbol{V}} \times \boldsymbol{m}) + (\boldsymbol{V} \times \dot{\boldsymbol{m}})] \times \boldsymbol{B} + \varepsilon_0(\boldsymbol{V} \times \boldsymbol{m}) \times (\boldsymbol{V} \cdot \boldsymbol{\nabla})\boldsymbol{B}. \tag{F10}$$

$$\int_{\text{all space}}(\mu_0^{-1}\boldsymbol{\nabla} \times \boldsymbol{\mathcal{M}}) \times \boldsymbol{B}dv = -\int_{-\infty}^{\infty}\boldsymbol{\nabla}(\boldsymbol{\mathcal{M}} \cdot \boldsymbol{H})dv^{\nearrow 0} + \int[(\boldsymbol{\mathcal{M}} \cdot \boldsymbol{\nabla})\boldsymbol{H} + (\boldsymbol{H} \cdot \boldsymbol{\nabla})\boldsymbol{\mathcal{M}} + \boldsymbol{\mathcal{M}} \times (\boldsymbol{\nabla} \times \boldsymbol{H})]dv$$

$$= (\boldsymbol{m} \cdot \boldsymbol{\nabla})\boldsymbol{H} - (\boldsymbol{\nabla} \cdot \boldsymbol{H})\boldsymbol{m}^{\nearrow 0} + \boldsymbol{m} \times \varepsilon_0 \partial \boldsymbol{E}/\partial t. \tag{F11}$$

---

We close this appendix by directly computing the Einstein-Laub force acting on a moving magnetic dipole (accompanied, as always, by the relativistically-induced electric dipole), showing that indeed it equals the Lorentz force minus the ubiquitous hidden momentum contribution.

$$\boldsymbol{F}_{\text{EL}}(t) = \int_{\text{all space}} \left\{ [\underbrace{(\varepsilon_0 \boldsymbol{V} \times \boldsymbol{\mathcal{M}})}_{\boldsymbol{\mathcal{P}}(\boldsymbol{r},t)} \cdot \boldsymbol{\nabla}]\boldsymbol{E} + \underbrace{[\partial(\varepsilon_0 \boldsymbol{V} \times \boldsymbol{\mathcal{M}})/\partial t]}_{\partial \boldsymbol{\mathcal{P}}(\boldsymbol{r},t)/\partial t} \times \mu_0 \boldsymbol{H} + (\boldsymbol{\mathcal{M}} \cdot \boldsymbol{\nabla})\boldsymbol{H} - \left(\frac{\partial \boldsymbol{\mathcal{M}}}{\partial t} \times \varepsilon_0 \boldsymbol{E}\right) \right\} dv. \tag{F12}$$

The first and third terms of the above integral are found to be

$$\int_{-\infty}^{\infty}(\boldsymbol{\mathcal{P}} \cdot \boldsymbol{\nabla})\boldsymbol{E}dv \underset{\text{Integration by parts}}{=} \iint_{-\infty}^{\infty}(\mathcal{P}_x \boldsymbol{E})|_{x=-\infty}^{\infty \, \nearrow 0}dydz + \iint_{-\infty}^{\infty}(\mathcal{P}_y \boldsymbol{E})\Big|_{y=-\infty}^{\infty \, \nearrow 0}dxdz + \iint_{-\infty}^{\infty}(\mathcal{P}_z \boldsymbol{E})|_{z=-\infty}^{\infty \, \nearrow 0}dxdy$$

$$-\int_{-\infty}^{\infty}(\boldsymbol{\nabla} \cdot \boldsymbol{\mathcal{P}})\boldsymbol{E}dv = \int_{-\infty}^{\infty}\rho(\boldsymbol{r},t)\boldsymbol{E}(\boldsymbol{r},t)dv. \tag{F13}$$

$$\int(\boldsymbol{\mathcal{M}} \cdot \boldsymbol{\nabla})\boldsymbol{H}dv = \int \boldsymbol{\nabla}(\boldsymbol{\mathcal{M}} \cdot \boldsymbol{H})dv - \int(\boldsymbol{H} \cdot \boldsymbol{\nabla})\boldsymbol{\mathcal{M}}dv - \int \boldsymbol{\mathcal{M}} \times (\boldsymbol{\nabla} \times \boldsymbol{H})dv - \int \boldsymbol{H} \times (\boldsymbol{\nabla} \times \boldsymbol{\mathcal{M}})dv.$$

[$\boldsymbol{\nabla}(\boldsymbol{a} \cdot \boldsymbol{b}) = (\boldsymbol{a} \cdot \boldsymbol{\nabla})\boldsymbol{b} + (\boldsymbol{b} \cdot \boldsymbol{\nabla})\boldsymbol{a} + \boldsymbol{a} \times (\boldsymbol{\nabla} \times \boldsymbol{b}) + \boldsymbol{b} \times (\boldsymbol{\nabla} \times \boldsymbol{a})$]   [$\varepsilon_0 \partial \boldsymbol{E}/\partial t$] (F14)

On the right-hand side of Eq.(F14), we have

$$\int_{\text{all space}} \boldsymbol{\nabla}(\boldsymbol{\mathcal{M}} \cdot \boldsymbol{H}) \mathrm{d}v = \iiint_{-\infty}^{\infty} [\partial_x(\boldsymbol{\mathcal{M}} \cdot \boldsymbol{H})\hat{\boldsymbol{x}} + \partial_y(\boldsymbol{\mathcal{M}} \cdot \boldsymbol{H})\hat{\boldsymbol{y}} + \partial_z(\boldsymbol{\mathcal{M}} \cdot \boldsymbol{H})\hat{\boldsymbol{z}}] \mathrm{d}x \mathrm{d}y \mathrm{d}z = 0. \quad (\text{F15})$$

$$\int_{\text{all space}} (\boldsymbol{H} \cdot \boldsymbol{\nabla}) \boldsymbol{\mathcal{M}} \, \mathrm{d}v \underset{\text{Integration by parts}}{=} -\int_{\text{all space}} (\boldsymbol{\nabla} \cdot \boldsymbol{H}) \boldsymbol{\mathcal{M}} \, \mathrm{d}v = 0. \quad [\boldsymbol{\nabla} \cdot \boldsymbol{B} = \mu_0 \boldsymbol{\nabla} \cdot \boldsymbol{H} = 0] \quad (\text{F16})$$

---

§A similar expression can be derived for the EM force acting on a moving electric dipole $\boldsymbol{p}(t)$ accompanied by the relativistically-induced magnetic dipole $\boldsymbol{m}(t)$. The equation of motion $\dot{\boldsymbol{p}} = (\boldsymbol{p} \cdot \boldsymbol{\nabla})\boldsymbol{E} + (\dot{\boldsymbol{p}} \times \mu_0 \boldsymbol{H}) + \boldsymbol{V} \times [(\boldsymbol{p} \cdot \boldsymbol{\nabla})\mu_0 \boldsymbol{H}]$ thus obtained has a natural physical interpretation in terms of the Lorentz force acting on the moving pair of equal and opposite electric charges of $\boldsymbol{p}(t)$. It is not difficult to prove that this equation is in complete accord with Eq.(15).

Substituting in Eq.(F12) from Eqs.(F13)-(F16), we finally arrive at

$$\boldsymbol{F}_{\text{EL}}(t) = \int_{\text{all space}} \left\{ \rho(\boldsymbol{r},t)\boldsymbol{E}(\boldsymbol{r},t) + \underbrace{\left(\frac{\partial \boldsymbol{\mathcal{P}}}{\partial t} + \mu_0^{-1} \boldsymbol{\nabla} \times \boldsymbol{\mathcal{M}}\right)}_{\boldsymbol{J}(\boldsymbol{r},t)} \times \mu_0 \boldsymbol{H}(\boldsymbol{r},t) - \frac{\partial}{\partial t}(\boldsymbol{\mathcal{M}} \times \varepsilon_0 \boldsymbol{E}) \right\} \mathrm{d}v. \quad (\text{F17})$$

Aside from the last term of the integrand, which is the time-rate of change of the hidden momentum density, Eq.(F17) is the expression of the Lorentz force acting on the electric charge and current densities associated with the moving dipole pair.

---

**Digression**: It is of some interest to relate the Einstein-Laub force, given by either Eq.(F12) or Eq.(F17), to the force derived from the $\boldsymbol{p} \cdot \boldsymbol{E} + \boldsymbol{m} \cdot \boldsymbol{H}$ Lagrangian given by Eq.(F8). The first term in the expression of $\boldsymbol{F}_{\text{EL}}$ as given in Eq.(F12) can be written [using straightforward integration, or invoking Eqs.(F13) and (F9)] as

$$\int_{\text{all space}} [(\varepsilon_0 \boldsymbol{V} \times \boldsymbol{\mathcal{M}}) \cdot \boldsymbol{\nabla}] \boldsymbol{E} \mathrm{d}v = [(\varepsilon_0 \boldsymbol{V} \times \boldsymbol{m}) \cdot \boldsymbol{\nabla}] \boldsymbol{E} = \varepsilon_0(\boldsymbol{V} \times \boldsymbol{m})(\boldsymbol{\nabla} \cdot \boldsymbol{E})^{\!\!\nearrow 0} - \varepsilon_0 \boldsymbol{\nabla} \times [(\boldsymbol{V} \times \boldsymbol{m}) \times \boldsymbol{E}]$$

$$= \varepsilon_0 \boldsymbol{\nabla} \times [(\boldsymbol{m} \cdot \boldsymbol{E})\boldsymbol{V}] - \varepsilon_0 \boldsymbol{\nabla} \times [(\boldsymbol{V} \cdot \boldsymbol{E})\boldsymbol{m}] = \varepsilon_0 [\boldsymbol{\nabla}(\boldsymbol{m} \cdot \boldsymbol{E})] \times \boldsymbol{V} - \varepsilon_0 \boldsymbol{\nabla} \times [(\boldsymbol{V} \cdot \boldsymbol{E})\boldsymbol{m}]$$

$$= \varepsilon_0 [(\boldsymbol{m} \cdot \boldsymbol{\nabla})\boldsymbol{E} + \boldsymbol{m} \times (\boldsymbol{\nabla} \times \boldsymbol{E})] \times \boldsymbol{V} - \varepsilon_0 \boldsymbol{\nabla} \times [(\boldsymbol{V} \cdot \boldsymbol{E})\boldsymbol{m}]$$

$$[\boldsymbol{\nabla} \times \boldsymbol{E} = -\partial \boldsymbol{B}/\partial t] \rightarrow = -\varepsilon_0 \boldsymbol{V} \times [(\boldsymbol{m} \cdot \boldsymbol{\nabla})\boldsymbol{E}] + \varepsilon_0 \boldsymbol{V} \times (\boldsymbol{m} \times \partial_t \boldsymbol{B}) - \varepsilon_0 \boldsymbol{\nabla} \times [(\boldsymbol{V} \cdot \boldsymbol{E})\boldsymbol{m}]$$

$$= -\varepsilon_0 \boldsymbol{V} \times [(\boldsymbol{m} \cdot \boldsymbol{\nabla})\boldsymbol{E}] + \varepsilon_0 (\boldsymbol{V} \cdot \partial_t \boldsymbol{B})\boldsymbol{m} - \varepsilon_0 (\boldsymbol{m} \cdot \boldsymbol{V})\partial_t \boldsymbol{B} - \varepsilon_0 \boldsymbol{\nabla} \times [(\boldsymbol{V} \cdot \boldsymbol{E})\boldsymbol{m}]. \quad (\text{F18})$$

Similarly, the last term of Eq.(F12) may be rearranged, as follows:

$$\int_{\text{all space}} (\partial \boldsymbol{\mathcal{M}}/\partial t) \times \varepsilon_0 \boldsymbol{E} \, \mathrm{d}v = \varepsilon_0 \dot{\boldsymbol{m}} \times \boldsymbol{E} + \varepsilon_0 \boldsymbol{m} \times (\boldsymbol{V} \cdot \boldsymbol{\nabla})\boldsymbol{E}$$

$$= \varepsilon_0 \dot{\boldsymbol{m}} \times \boldsymbol{E} + \varepsilon_0 \boldsymbol{m} \times [\boldsymbol{\nabla}(\boldsymbol{V} \cdot \boldsymbol{E}) - \boldsymbol{V} \times (\boldsymbol{\nabla} \times \boldsymbol{E})] \leftarrow [\boldsymbol{\nabla} \times \boldsymbol{E} = -\partial \boldsymbol{B}/\partial t]$$

$$= \varepsilon_0 \dot{\boldsymbol{m}} \times \boldsymbol{E} + \underbrace{\varepsilon_0 \boldsymbol{m} \times [\boldsymbol{\nabla}(\boldsymbol{V} \cdot \boldsymbol{E})]}_{\boldsymbol{\nabla} \times (\psi \boldsymbol{a}) = \boldsymbol{\nabla}\psi \times \boldsymbol{a} + \psi \boldsymbol{\nabla} \times \boldsymbol{a}} + \varepsilon_0 \boldsymbol{m} \times (\boldsymbol{V} \times \partial_t \boldsymbol{B})$$

$$= \varepsilon_0 \dot{\boldsymbol{m}} \times \boldsymbol{E} - \varepsilon_0 \boldsymbol{\nabla} \times [(\boldsymbol{V} \cdot \boldsymbol{E})\boldsymbol{m}] + \varepsilon_0 (\boldsymbol{m} \cdot \partial_t \boldsymbol{B})\boldsymbol{V} - \varepsilon_0 (\boldsymbol{m} \cdot \boldsymbol{V})\partial_t \boldsymbol{B}. \quad (\text{F19})$$

Combining Eqs.(F18) and (F19), we will have

$$\int_{\text{all space}} \{[(\varepsilon_0 \boldsymbol{V} \times \boldsymbol{\mathcal{M}}) \cdot \boldsymbol{\nabla}]\boldsymbol{E} - (\partial \boldsymbol{\mathcal{M}}/\partial t) \times \varepsilon_0 \boldsymbol{E}\} \mathrm{d}v = -\varepsilon_0 \dot{\boldsymbol{m}} \times \boldsymbol{E} - \varepsilon_0 \boldsymbol{V} \times [(\boldsymbol{m} \cdot \boldsymbol{\nabla})\boldsymbol{E}] + (\varepsilon_0 \boldsymbol{V} \times \boldsymbol{m}) \times \partial_t \boldsymbol{B}. \quad (\text{F20})$$

Substituting for the first and last terms of Eq.(F12) from Eq.(F20), we finally arrive at

$$\boldsymbol{F}_{\text{EL}}(t) = (\boldsymbol{m} \cdot \boldsymbol{\nabla})\boldsymbol{H} - \varepsilon_0 \dot{\boldsymbol{m}} \times \boldsymbol{E} - \varepsilon_0 \boldsymbol{V} \times [(\boldsymbol{m} \cdot \boldsymbol{\nabla})\boldsymbol{E}] + \frac{\mathrm{d}}{\mathrm{d}t}[\boldsymbol{p}(t) \times \mu_0 \boldsymbol{H}(\boldsymbol{r}_p, t)]. \quad (\text{F21})$$

Comparing Eq.(F21) with Eq.(F8), we find that the Einstein-Laub force acting on the moving magnetic dipole differs from that obtained via the $\boldsymbol{p} \cdot \boldsymbol{E} + \boldsymbol{m} \cdot \boldsymbol{H}$ Lagrangian by $\mathrm{d}[\boldsymbol{p}(t) \times \mu_0 \boldsymbol{H}(\boldsymbol{r}_p, t)]/\mathrm{d}t$.

---